\documentclass[aps,prl,reprint,amsmath,amssymb,groupedaddress]{revtex4-2}
\usepackage[T1]{fontenc}
\usepackage{lmodern}
\usepackage{mathtools,bm}
\usepackage{booktabs,array,longtable,graphicx,xcolor}
\usepackage{microtype}
\usepackage{hyperref}
\hypersetup{colorlinks=true,linkcolor=blue!45!black,citecolor=blue!45!black,
 urlcolor=blue!45!black,pdftitle={A full-color four-point form factor},
 pdfauthor={Song He and Xuhang Jiang},pdfsubject={PRL manuscript and Supplemental Material}}
\newif\ifsupplementary
\supplementarytrue
\newcommand{\tr}{\operatorname{tr}}
\newcommand{\PT}{\operatorname{PT}}

\newcommand{\LP}{\operatorname{LP}}
\newcommand{\FF}{\mathcal F}
\newcommand{\HH}{\mathcal H}
\newcommand{\Symb}{\mathcal S}
\newcommand{\Ctwotwo}{\mathcal C_{22}}
\newcommand{\Q}{\mathbb Q}
\newcommand{\aset}{\mathcal A}
\newcommand{\eps}{\epsilon}
\newcommand{\sourcefile}[1]{\nolinkurl{#1}}
\newcommand{\prlsection}[1]{\par\medskip\indent\textit{#1.---}\ignorespaces}
\begin{document}
\title{A full-color four-point form factor}
\author{Song He}
\author{Xuhang Jiang}
\affiliation{New Cornerstone Laboratory, Institute of Theoretical Physics,
Chinese Academy of Sciences,
Beijing 100190, China}
\begin{abstract}
We determine the symbol and complete functions of the full-color two-loop four-point
maximally helicity violating stress-tensor form factor in $\mathcal N=4$
super Yang--Mills theory, including $N^1$ double traces and $N^0$ single traces.
Numerical unitarity and a master-integral bootstrap fix the result.
A direct form-factor bootstrap reproduces the symbol.
Permuting external legs in the planar 88-letter alphabet yields all 79 letters,
while the $N^0$ terms contain four new nonclassical contributions.
We also obtain the complete $N^0$ two-loop triple-collinear hard splitting functions.
\end{abstract}
\maketitle

\prlsection{Introduction}
Form factors describe the coupling of a local operator to on-shell
particles. They probe gauge-theory dynamics with an off-shell momentum
insertion. For the stress-tensor multiplet in $\mathcal N=4$
super Yang--Mills theory, the planar four-point maximally helicity
violating (MHV) form factor is known at two loops, including all terms
beyond the symbol~\cite{planar-bootstrap,planar,planar-function}.
The subleading-color contribution has been missing. Its calculation
tests how much of the planar analytic structure survives at finite $N$,
where nonplanar integrals can introduce new functions.
Full-color two-loop five-point amplitude symbols provide a related
example~\cite{fivepoint}.

We determine the full-color symbol and complete functions
using numerical generalized unitarity and a master-integral bootstrap.
We also perform a direct form-factor bootstrap in a larger symbol space,
using the same unitarity data.
All letters follow from permutations of the external particles
in the planar alphabet, but the new single-trace coefficients contain
four nonclassical contributions outside the span of the known crossed
planar MHV functions~\cite{planar-bootstrap,planar,planar-function} and
next-to-MHV (NMHV) functions~\cite{nmhv}.
We also obtain the complete $N^0$ two-loop triple-collinear hard splitting
functions. Their scale logarithms are explicit, and the four finite parts
follow from one $26$-letter function and its permutations.
The expressions, derivations and numerical checks are supplied in the
Supplemental Material \cite{supplemental}.

\prlsection{Kinematics and color decomposition}
We consider the MHV form factor of the half-BPS operator
$\mathcal O=\tr(\phi^2)$ in $SU(N)$ gauge theory,
\begin{equation}
 \FF_4(1,2,3,4;q)=\langle1,2,3,4|\mathcal O|0\rangle,
 \qquad q=\sum_{i=1}^4p_i,
 \label{eq:object}
\end{equation}
with $p_i^2=0$ and $q^2\ne0$.
Writing $s_I=(\sum_{i\in I}p_i)^2$, we choose the five independent
dimensionless variables $u_i=s_{i,i+1}/q^2$ for $i=1,2,3$ and
$v_i=s_{i,i+1,i+2}/q^2$ for $i=1,2$.
The infrared-finite form factor is
\begin{equation}
 |\HH\rangle=\lim_{\eps\to0}\mathbf Z^{-1}|\FF\rangle,
 \label{eq:hard}
\end{equation}
where $\mathbf Z$ contains only poles in dimensional reduction,
$D=4-2\eps$, and is fixed through two loops by the color-dipole anomalous
dimension \cite{infrared}. We give the result at $\mu^2=q^2$ and use
$q^2=1$ as the unit of mass squared.

We suppress the tree coupling and the MHV numerator, which is
$\langle rs\rangle^2$ for two scalar legs $r,s$ and positive-helicity
gluons on the remaining legs \cite{one-loop}. The kinematic coefficients
use two independent Parke--Taylor factors,
$P_1=\PT(1234)$ and $P_2=\PT(1243)$, with
$\PT(ijkl)=1/(\langle ij\rangle\langle jk\rangle
\langle kl\rangle\langle li\rangle)$.
The third ordering obeys $P_3=\PT(1324)=-P_1-P_2$.
We use fundamental generators $t^a$ with
$\tr(t^at^b)=\delta^{ab}/2$ and write
$\tr(ijkl)=\tr(t^{a_i}t^{a_j}t^{a_k}t^{a_l})$,
where $a_i$ is the adjoint color index of external leg $i$.
The tensors $S_i$ are a single trace plus its reverse, with orderings
$(1234,1243,1324)$, and $D_i$ are products of two traces with pairings
$(12|34,13|24,14|23)$.
For example, $S_1=\tr(1234)+\tr(1432)$ and
$D_1=\tr(12)\tr(34)$.

The two-loop hard function takes the form \cite{prefactor-note}:
\begin{equation}
\begin{aligned}
 \HH_4^{(2)}
 ={}&\sum_{i=1}^3\Bigl[S_i\bigl(N^2L_i+C_i\bigr)+2NKD_i\Bigr]\\
 &-N\bigl(D_1C_3+D_2C_1+D_3C_2\bigr).
\end{aligned}
\label{eq:mainresult}
\end{equation}
The expansion and normalization are specified in the Supplemental
Material.
The $L_i$ are planar coefficients. The $N^1$ double traces and
$N^0$ single traces are parametrized by six weight-four functions,
\begin{equation}
\begin{aligned}
 C_i&=P_1f_{i1}+P_2f_{i2}\quad(i=1,2),\\
 C_3&=-C_1-C_2,\qquad K=P_1k_1+P_2k_2.
\end{aligned}
\label{eq:scalarrows}
\end{equation}
The color relations \cite{color-original,color} imply $K=L_1+L_2+L_3$.
Writing $L_i=P_i h_i$ gives
\begin{equation}
 k_1=h_1-h_3,\qquad k_2=h_2-h_3.
 \label{eq:planarK}
\end{equation}
These relations require the same infrared subtraction and analytic
continuation for the planar and subleading-color functions.
We determine $K$ together with the $C_i$, without imposing its planar
value.

\begin{samepage}
Crossing relates the four $N^0$ functions:
\begin{equation}
 f_{12}=(1-\tau)f_{11},\quad
 f_{21}=\sigma f_{12},\quad f_{22}=\sigma f_{11}.
 \label{eq:seed}
\end{equation}
Here $\tau=(13)$ and $\sigma=(12)$ are permutations acting on the
external particles. Their action includes the sign change of the
oriented parity-odd invariant and the corresponding analytic
continuation of the functions.
\end{samepage}
Together with the planar hard functions reconstructed below, the complete seed
$f_{11}$ therefore determines the full two-loop color decomposition.

\prlsection{Numerical unitarity and master-integral bootstrap}
Generalized unitarity determines the cuts from a tree form factor and
tree amplitudes,
\begin{equation}
 \operatorname{Cut}_{\mathcal C}\FF_4^{(2)}
 =\sum_{\text{states, colors}}\FF_{\mathcal O}^{(0)}
   \prod_v\mathcal A_v^{(0)}.
 \label{eq:unitarity}
\end{equation}
We evaluate the rational cut coefficients over finite fields
\cite{unitarity,finitefields} and apply integration-by-parts reductions
at the sampled kinematics. The reduced integrals are expressed in the
known one-mass pentagon functions \cite{integrals}.
Our implementation uses \texttt{Singular}~\cite{singular} for cut-compatible
relations, \texttt{Kira} and \texttt{Fermat}~\cite{kira,fermat} for selected uncut reductions,
and \texttt{FLINT}~\cite{flint} for finite-field linear algebra.
\texttt{ColorMath}~\cite{colormath} provides independent color-algebra checks.
The calculation retains the $D$-dimensional numerator terms and the
one-loop form factor through $\eps^2$. We perform the color-space
subtraction in Eq.~\eqref{eq:hard} before extracting the symbol.

In the master-integral bootstrap~\cite{mi-bootstrap}, the unknowns are the coefficients of
the finite form factor in the space generated by the integral functions
and their products. Denote its weight-four symbol basis by
$B^{\rm MI}_\alpha$. For
$g_r\in\{f_{11},f_{12},f_{21},f_{22},k_1,k_2\}$ we write
\begin{equation}
 \Symb(g_r)=\sum_{\alpha=1}^{6468}c_{r\alpha}B^{\rm MI}_\alpha,
 \qquad c_{r\alpha}\in\Q.
 \label{eq:mi-ansatz}
\end{equation}
Integrability and the physical first entries are built into this basis.
Crossing symmetry acts jointly on color tensors, rational prefactors
and scalar functions. It reduces the $38808$ coefficients to $1767$.
Numerical generalized unitarity fixes all $1767$ parameters in this
ansatz without timelike collinear, soft or ordinary Steinmann input.
Imposing these conditions first leaves $641$, $535$ and $267$
parameters, respectively.

At each finite-field point, integral reduction and infrared subtraction
give a vector of rational coefficients multiplying formal transcendental
basis elements. Crossing separates the two Parke--Taylor coefficients
after projecting this vector to the symbol.
We reconstruct the rational coefficients from two primes and check them
at an additional prime and further kinematic points.
The vectors before this projection are retained for the
complete-function matching below.

\prlsection{Direct form-factor bootstrap}
For the direct form-factor bootstrap, we construct the symbol ansatz
from the one-mass five-point alphabet, without restricting it to the
master-integral span.
At fixed $q^2$, the alphabet has $203$ independent letters, where
independence means $\Q$-linear independence of their logarithmic
differentials. The omitted scale letter is $W_1=q^2$.
We allow the six two-particle and four three-particle invariants as
first entries and impose integrability in every adjacent pair of slots.
The resulting spaces $\mathcal V_w$ have dimensions
\begin{equation}
 (\dim\mathcal V_1,\ldots,\dim\mathcal V_4)
 =(10,108,1149,11997),
 \label{eq:spaces}
\end{equation}
compared with $(10,90,787,6468)$ in the integral-generated space.
The initial master-integral and direct ans\"atze for the six scalar functions contain
$6\times6468=38808$ and $6\times11997=71982$ coefficients,
respectively, before imposing crossing, as listed in Table~\ref{tab:bootstrap}.

Crossing leaves $3135$ parameters, all fixed by the same numerical
unitarity data. We also test timelike collinear and soft factorization.
The soft limit has a nonzero right-hand side, obtained from the
soft current $J$ \cite{soft} and lower-point form factors in the same
subtraction scheme. Its finite operator is
\begin{equation}
 S_{\rm hard}=\lim_{\eps\to0}\mathbf Z_4^{-1}J\mathbf Z_3.
 \label{eq:softtransfer}
\end{equation}
The symbols obey ordinary Steinmann relations for the twelve ordered
pairs of distinct overlapping three-particle channels.
No last-entry conditions, extended Steinmann relations or restriction
to the final $79$ letters are imposed.

\begin{table}[t]
\centering
\begin{tabular}{@{}lrr@{}}
\toprule
Stage & Master-integral & Direct\\\midrule
Initial ansatz &38808&71982\\
Crossing &1767&3135\\
Collinear &641&1067\\
Soft &535&801\\
Steinmann &267&431\\
Numerical unitarity &0&0\\\bottomrule
\end{tabular}
\caption{Undetermined symbol coefficients after successive
constraints in the master-integral and direct form-factor ans\"atze,
retaining the nonzero soft right-hand side. The complete numerical-unitarity symbol-coefficient
data can also fix all $1767$ and $3135$ parameters, respectively, before
collinear, soft and ordinary Steinmann constraints.}
\label{tab:bootstrap}
\end{table}

The direct form-factor bootstrap reproduces the symbols obtained in
the master-integral bootstrap. The $79$-letter alphabet is therefore
obtained from the larger $11997$-dimensional integrable space, without
restricting the ansatz to the $6468$-dimensional master-integral span.

\begin{table*}[t]
\centering
\setlength{\tabcolsep}{6pt}
\begin{tabular}{@{}llrrrrrr@{}}
\toprule
Point & Part & $f_{11}$ & $f_{12}$ & $f_{21}$ & $f_{22}$ & $k_1$ & $k_2$ \\
\midrule
$X_0$ & $\operatorname{Re}$ & 425.88946491 & 642.90479304 & -392.54707932 & -292.46170557 & 21.65142219 & 21.65142219 \\
 & $\operatorname{Im}$ & 199.03959834 & 346.79465047 & 366.52075369 & 82.05892865 & 44.55608713 & 44.55608713 \\
\midrule
$X_1$ & $\operatorname{Re}$ & 419.71401634 & 647.29738355 & -414.43023451 & -301.37695752 & 5.56909238 & 1.27994501 \\
 & $\operatorname{Im}$ & 205.14658779 & 376.71829923 & 397.35430159 & 101.85251505 & 70.06932746 & 73.93066164 \\
\bottomrule
\end{tabular}
\caption{Complete hard-function coefficients in the minimal-pole scheme
at Eq.~\eqref{eq:numerical-points}, rounded to eight decimal places.
Both points lie in our $s_{34}$ production channel.}
\label{tab:full-function-benchmark}
\end{table*}

\prlsection{Complete functions}
Starting from the reconstructed symbols, we expand all six subleading-color functions
in the pentagon-function basis of Ref.~\cite{integrals}, including its
boundary constants.
Each scalar function has the form
\begin{equation}
\begin{aligned}
 g={}&g^{[4]}+I g^{[3]}+I^2 g^{[2]}+I^3 g^{[1]}\\
 &+\zeta_3\widetilde g^{[1]}+\alpha I^4+\beta I\zeta_3,
 \qquad I=i\pi,
\end{aligned}
\label{eq:uplift}
\end{equation}
where a bracketed superscript denotes the total weight of a rational
linear combination of basis functions and their products, and
$\alpha,\beta\in\Q$.
The terms other than $g^{[4]}$ are invisible to the ordinary symbol.
Matching to the numerical unitarity calculation in the full function
basis, before taking the symbol, fixes all remaining coefficients.
The complete expressions
are provided in the ancillary files.

The functions are specified in the $s_{34}$ production channel,
$(-p_3)+(-p_4)\to(-q)+p_1+p_2$.
With $(\widehat p_1,\ldots,\widehat p_5)=(-q,p_1,p_2,p_3,p_4)$,
this is the $s_{45}$ channel of Ref.~\cite{integrals}.
We denote its invariants by $S_{ij}=(\widehat p_i+\widehat p_j)^2$,
so $S_{45}=s_{34}=q^2u_3$.
We use the reference's principal square root and choose
$\tr_5/\sqrt{\Delta_5}=+1$ at the reference point, where $\tr_5$ is the oriented parity-odd
invariant and $\Delta_5=(\tr_5)^2$.

Within the pure-function ansatz, we also reconstruct the three leading-color
coefficients $L_i=P_i h_i$ independently from numerical unitarity,
without using the known planar
results~\cite{planar-bootstrap,planar,planar-function} as input.
After converting to the remainder convention of Ref.~\cite{planar},
all three agree exactly with its complete planar expression and its
crossings, including every term beyond the symbol.
The comparison is described in the Supplemental Material. The ancillary
data include the pure functions $h_i=L_i/P_i$ and their symbols,
with Parke--Taylor factors supplied separately.

Table~\ref{tab:full-function-benchmark} gives all six complete subleading-color functions
at $q^2=\mu^2=1$. In our variables, the two production points are
\begin{equation}
\begin{aligned}
 (u_1,u_2,u_3,v_1,v_2)\big|_{X_0}&=(2,-2,7,-2,3),\\
 (u_1,u_2,u_3,v_1,v_2)\big|_{X_1}&=(2.23,-1.87,7.19,-1.73,3.1),
\end{aligned}
\label{eq:numerical-points}
\end{equation}
where the displayed decimals are exact and
$\tr_5/\sqrt{\Delta_5}=+1$ at both points.
At $X_0$, we contract the exact coefficient polynomials with the published
boundary values~\cite{integrals}, using high-precision arithmetic in
\texttt{mpmath}~\cite{mpmath} and an independent \texttt{Mathematica} evaluation~\cite{mathematica}.
At $X_1$, we use \texttt{PentagonFunctions++}~\cite{integrals,pentagon-cpp}
with double-double and quad-double arithmetic from \texttt{QD}~\cite{qd}.
All terms beyond the symbol in Eq.~\eqref{eq:uplift} are included.
The largest absolute DD--QD difference is $5.62\times10^{-24}$.
Evaluations with \texttt{PentagonFunctions++} converge to the reference
values as the kinematics approach $X_0$.
Precision checks and a comparison in the decay channel are included
in the Supplemental Material.

Crossing is verified exactly in this representation.
At a new kinematic point over an independent prime field, the coefficients of the
function basis agree with the unitarity calculation and all infrared
poles vanish. Two further generalized cuts agree at that kinematic point,
without refitting.

We verify soft and collinear factorization numerically, including the
finite boundary coefficients, by transporting the complete functions
to the boundary. These coefficients are extracted without imposing
their expected values. Numerical double discontinuities also agree
with Steinmann relations for the common continuation paths specified
in the Supplemental Material.

Away from other singularities, all six functions remain finite as the
Gram determinant $\Delta_5$ approaches zero. Near a generic configuration
with $\Delta_5=0$, each function has the expansion
\begin{equation}
 g=g_0+g_1\sqrt{\Delta_5}+g_2\Delta_5+\cdots,
 \label{eq:gram}
\end{equation}
where the coefficients depend on the remaining kinematic variables.
There are no logarithms or negative powers of $\sqrt{\Delta_5}$,
including in the terms beyond the symbol. We present in the Supplemental Material its derivation and
conditions.

\prlsection{Alphabet and new nonclassical functions}
The six complete functions, including all terms beyond the symbol,
use the same $79$-letter alphabet as their symbols,
\begin{equation}
 \aset_{79}=\{W_{2:33},\,W_{118:137},\,W_{162:188}\},
 \label{eq:79}
\end{equation}
where $W_{a:b}$ denotes an inclusive range in the alphabet of
Ref.~\cite{integrals}. Their logarithmic span is closed under $S_4$.
The other $124$ fixed-scale letters do not appear.

Let $V(\aset)=\operatorname{span}_{\Q}\{d\log a:a\in\aset\}$ and
$V_{S_4}(\aset)=\sum_{\pi\in S_4}\pi^*V(\aset)$.
The planar two-loop MHV alphabet has $34$ independent letters
\cite{planar-bootstrap}. For this alphabet and the larger planar
$88$-letter alphabet discussed in Ref.~\cite{nmhv}, we find
\begin{equation}
 \dim V_{S_4}(\aset_{34})=61,\qquad
 \dim V_{S_4}(\aset_{88})=166.
 \label{eq:closure}
\end{equation}
Every letter of $\aset_{79}$ is a permuted planar letter, up to
inversion and a constant:
\begin{equation}
 d\log W_j=\eta_j\,d\log a_{r_j}(\pi_j\cdot p),
 \qquad \eta_j=\pm1.
 \label{eq:lettermap}
\end{equation}
Here $a_r$ are the letters of $\aset_{88}$ and $\pi_j\in S_4$
permutes the external momenta.
For example, $d\log W_{17}=d\log a_{18}(1243\cdot p)$ and
$d\log W_{162}=-d\log a_{69}(1243\cdot p)$.
Hence $V(\aset_{79})\subset V_{S_4}(\aset_{88})$.
For the original ordering of $\aset_{88}$,
$\dim[V(\aset_{79})\cap V(\aset_{88})]=50$.

\begin{table}[t]
\centering
\begin{tabular}{@{}lrr@{}}
\toprule
Functions & $W$ letters used & $\dim L$\\\midrule
$f_{11},f_{12},f_{21},f_{22}$ &64&64\\
$k_1,k_2$ &67&61\\
All six &79&79\\\bottomrule
\end{tabular}
\caption{Symbol support and the dimension of the smallest common
letter space $L$ for which the symbols lie in $L^{\otimes4}$.
The $k_i$ symbols use $67$ $W$ letters but depend on
$61$ linear combinations of their logarithmic differentials.}
\label{tab:letterspaces}
\end{table}

Both $k_i$ symbols lie in $V_{S_4}(\aset_{34})^{\otimes4}$, as expected
from Eq.~\eqref{eq:planarK}. The symbol of each $f_{ij}$ lies outside this space.
Table~\ref{tab:letterspaces} distinguishes the support in the $W$ basis
from the smallest common letter space.
The full $79$-dimensional space extends the crossed planar MHV space
by $18$ independent letters: twelve rational and six algebraic.

To compare the functions with the known planar results, we apply
\begin{equation}
 \Ctwotwo(a\otimes b\otimes c\otimes d)
 =(a\wedge b)\wedge(c\wedge d)
 \label{eq:c22}
\end{equation}
to the weight-four symbols. This map annihilates classical
$\operatorname{Li}_4$ symbols and shuffle products \cite{classical-symbols}.
Let $\mathcal P$ consist of the five planar two-loop NMHV coefficient
functions \cite{nmhv} under all $24$ permutations, together with the
three ordered planar MHV hard functions. At fixed external labels,
\begin{equation}
\begin{aligned}
 \dim\operatorname{span}_{\Q}\Ctwotwo\Symb(\mathcal P)&=30,\\
 \dim\operatorname{span}_{\Q}\Ctwotwo\Symb(\mathcal P,f_{11},f_{12},f_{21},f_{22})&=34.
\end{aligned}
\label{eq:newdirections}
\end{equation}
We retain the full permuted alphabets in this comparison.
The four $N^0$ functions therefore give four independent nonclassical
contributions beyond these planar functions, even after allowing classical polylogarithms
and products. This independence is at fixed external labels. Crossing
in Eq.~\eqref{eq:seed} also permutes the kinematic arguments.

\prlsection{Triple-collinear splitting functions}
In the triple-collinear limit $s_{123}/q^2=\rho\to0$, the four-point
form factor factorizes into splitting currents and the two-point
Sudakov form factor. We determine the four pure functions entering
the $N^0$ two-loop MHV hard splitting current, including all scale
logarithms and finite parts.
Write $x=s_{12}/s_{123}$ and $y=s_{23}/s_{123}$. The longitudinal momentum
fractions are $z_i=(p_i\cdot p_4)/[(p_1+p_2+p_3)\cdot p_4]$,
with $z_1+z_2+z_3=1$. Set $h=1-x-y$.
For $q^2=\mu^2>0$, set
\begin{equation}
 L=\log\rho-i\pi,\qquad
 U=\log\frac{z_1y}{z_2h},\qquad V=\log\frac{z_3x}{z_2h}.
 \label{eq:triple-logs}
\end{equation}
The $N^0$ coefficients receive no contribution from lower-loop splitting
currents multiplying the one- or two-loop Sudakov form factor.
Color evolution and the known one-loop splitting amplitude
\cite{splitting-one-loop} fix all their scale logarithms.
At leading power, their $L^4$ and $L^3$ coefficients vanish. In particular,
\begin{equation}
 \begin{aligned}
 f_{11}={}&V(U+V)L^2-2(U+V)\kappa_1 L\\
          &+G_{11}(x,y,z_1,z_3)+o(1),
 \end{aligned}
 \label{eq:triple-f11}
\end{equation}
where $\kappa_1$ is a difference of weight-two one-loop functions.
Its explicit expression and the corresponding expansions of the
other three coefficients are given in the Supplemental Material.
Since $\tau$ and $\sigma$ leave $\rho$ unchanged, Eq.~\eqref{eq:seed}
also holds with $f_{ij}$ replaced by $G_{ij}$, with the same analytic
continuation prescription.

The four finite coefficients $G_{ij}$ each require $26$ independent
letters, and their smallest common alphabet has $29$.
The upper bounds follow from
the complete Chen representations, including the terms multiplying
transcendental constants. Matching lower bounds follow from their
nonclassical symbols.

The decay-channel pentagon functions
\cite{decay-functions,decay-data} give $G_{11}$ as an iterated integral
over its $26$ letters. To fix its boundary constants, we integrate in
$\rho=s_{123}/q^2$ from $1/2$ toward $0$ at fixed
$x=y=1/3$ and $z_1=z_2=z_3=1/3$.
The finite endpoint parts of these one-dimensional integrals are obtained
from Goncharov polylogarithms after two separate changes of variable.
The ancillary files specify the complete triple-collinear functions and
their symbols, including the boundary constants and letter definitions.
We compare their values with the triple-collinear limit
of the original four-point functions in the Supplemental Material.
The benchmarks in the Supplemental Material further show that
$G_{11}$ is not a function of the three hexagon cross ratios alone.

\prlsection{Discussion}
We have determined the full-color two-loop four-point MHV hard form factor,
including the $N^1$ double-trace and $N^0$ single-trace
coefficients and all terms beyond their symbols. The complete functions
use $79$ letters obtained by crossing the planar alphabet. Nevertheless,
the $N^0$ coefficients contain four independent nonclassical contributions
outside the span of the known crossed planar MHV and NMHV functions.
We also obtain the complete $N^0$ two-loop triple-collinear hard
splitting functions. Their scale logarithms are explicit, and their
four finite parts follow from one $26$-letter function and its permutations.
Restoring rational prefactors and color tensors gives the corresponding
splitting contribution contracted with the two-point tree form factor.

Within the stated two-Parke--Taylor ansatz, the same
numerical-unitarity data fix all symbol coefficients in the master-integral
and direct form-factor bootstraps after physical first entries,
integrability and crossing are imposed. Collinear, soft and ordinary
Steinmann relations provide further checks. The direct form-factor
bootstrap yields identical symbols from a larger ansatz.
Separate matching in the complete function basis determines all terms
beyond the symbol.

The known maximal-transcendentality relation between the two-loop
three-point remainder and the corresponding Higgs plus three-gluon
amplitude \cite{higgs}
motivates a comparison with Higgs plus four-parton amplitudes at
weight four in the heavy-top effective theory, with matched operator
component and infrared subtraction. The related three-loop three-point
form factor is also known analytically \cite{threepoint}.
A bootstrap construction and simplifications of the three-loop
three-point form factor are presented separately~\cite{threepoint-bootstrap}.

\begin{acknowledgments}
We thank Gang Yang for earlier discussions and Westlake University for hospitality.
GPT-6 Astra (OpenAI) assisted with manuscript preparation. Its use in the
alphabet and triple-collinear analyses is described in the Supplemental
Material. The authors take responsibility for the scientific content.
This work is supported by the National Natural Science Foundation of
China under Grant Nos.~12225510 and 12247103, and by the New Cornerstone
Science Foundation.
\end{acknowledgments}

\prlsection{Data availability}
The electronic files accompanying the Supplemental Material
\cite{supplemental} include \texttt{ff4\_ancillary.zip}, which contains
the nine pure functions and their symbols, the Parke--Taylor factors
and color tensors, and the triple-collinear limits of all nine functions
and their symbols. The archive also supplies the planar-letter map
and data for the nonclassical-function and triple-collinear letter-space
comparisons. The file \texttt{anc/README.wl} explains the conventions
and use of the analytic expressions; \texttt{anc/structural/README.TXT}
describes the comparison data and their checks.
The source pentagon functions are available in
Refs.~\cite{integrals,decay-data}.

\ifsupplementary
\onecolumngrid\clearpage
\small
\appendix
\setcounter{figure}{0}\setcounter{table}{0}\setcounter{equation}{0}
\renewcommand{\thefigure}{S\arabic{figure}}
\renewcommand{\thetable}{S\arabic{table}}
\renewcommand{\theequation}{S\arabic{equation}}
\renewcommand{\theHequation}{supp.\arabic{equation}}
\renewcommand{\theHfigure}{supp.\arabic{figure}}
\renewcommand{\theHtable}{supp.\arabic{table}}
\section*{Supplemental Material}
\begin{center}
\textbf{A full-color four-point form factor}\\[0.5em]
Song He and Xuhang Jiang\\[0.3em]
\textit{New Cornerstone Laboratory, Institute of Theoretical Physics,\\
Chinese Academy of Sciences, Beijing 100190, China}
\end{center}

\subsection*{I. Calculation and consistency checks}

\subsubsection*{Conventions and infrared subtraction}

We use the expansion and loop measure
\begin{equation}
 \FF=\sum_{\ell\ge0}a^\ell\FF^{(\ell)},\qquad
 a=\frac{g_{\rm YM}^2}{16\pi^2}(4\pi e^{-\gamma_E})^\eps,
 \qquad
 \int_\ell=e^{\gamma_E\eps}\mu^{2\eps}
          \int\frac{d^{4-2\eps}\ell}{i\pi^{2-\eps}}.
 \label{eq:coupling-measure}
\end{equation}
Here $g_{\rm YM}$ is dimensionless and the overall tree coupling is
removed. With the component numerator $\langle rs\rangle^2$ stripped
as in Ref.~\cite{one-loop}, the tree normalization is
$\FF^{(0)}=\sum_i S_iP_i$.
The product $a\int_\ell$ equals
$g_{\rm YM}^2\mu^{2\eps}\int d^{4-2\eps}\ell/[i(2\pi)^{4-2\eps}]$.
The integrated one-loop functions contain
\begin{equation}
 c_\Gamma=e^{\gamma_E\eps}
 \frac{\Gamma(1+\eps)\Gamma(1-\eps)^2}{\Gamma(1-2\eps)}
 =1-\tfrac12\zeta_2\eps^2+O(\eps^3).
\end{equation}
For the four external adjoint states, let
\begin{equation}
 L_{ij}=\log\frac{-s_{ij}-i0}{\mu^2},\qquad
 \mathbf M=-2\sum_{i<j}\mathbf T_i\!\cdot\!\mathbf T_j L_{ij},
\end{equation}
where $\mathbf T_i$ is the color generator acting on leg $i$.
The minimal-pole operator in dimensional reduction is
\begin{equation}
 \log\mathbf Z=
 a\left[-\frac{4N}{\eps^2}\mathbf1+\frac{\mathbf M}{\eps}\right]
 +a^2\left[\frac{2N^2\zeta_2}{\eps^2}\mathbf1
          -\frac{N\zeta_2}{\eps}\mathbf M
          +\frac{2N^2\zeta_3}{\eps}\mathbf1\right]+O(a^3).
 \label{eq:logZ}
\end{equation}
It follows from the two-loop dipole anomalous dimension
\cite{infrared}, and its planar poles agree with the normalization
of Ref.~\cite{planar}.
Expanding Eq.~\eqref{eq:hard} gives
\begin{equation}
 \HH^{(2)}=\lim_{\eps\to0}\left[
 \FF^{(2)}-\mathbf Z^{(1)}\FF^{(1)}
 +\bigl((\mathbf Z^{(1)})^2-\mathbf Z^{(2)}\bigr)\FF^{(0)}\right].
 \label{eq:subtraction2}
\end{equation}
In particular, the double pole of $\mathbf Z^{(1)}$ requires
$\FF^{(1)}$ through $\eps^2$.

We define $\tr(ijkl)=\tr(t^{a_i}t^{a_j}t^{a_k}t^{a_l})$.
The subleading-color contribution in Eq.~\eqref{eq:mainresult} is
\begin{align}
 \Gamma_1&=S_1-S_3+N(D_1-D_2),\nonumber\\
 \Gamma_2&=S_2-S_3+N(D_1-D_3),\qquad
 \Gamma_K=2N(D_1+D_2+D_3),\nonumber\\
 \HH_{\rm SLC}^{(2)}
 &=P_1(\Gamma_1f_{11}+\Gamma_2f_{21}+\Gamma_K k_1)
  +P_2(\Gamma_1f_{12}+\Gamma_2f_{22}+\Gamma_K k_2).
 \label{eq:compactgamma}
\end{align}
The double-trace ordering $(12|34,13|24,14|23)$ corresponds to
$N(2K-C_3,2K-C_1,2K-C_2)$.
The bivalent color vertex of $\tr(\phi^2)$ reduces to $\delta^{ab}$.
Removing it leaves the usual four-adjoint-leg color graphs.
Their group-theory relations give Eqs.~\eqref{eq:mainresult} and
\eqref{eq:planarK} \cite{color-original,color}.

For the integral variables, use
$(\widehat p_1,\ldots,\widehat p_5)=(-q,p_1,p_2,p_3,p_4)$,
set $S_{ij}=(\widehat p_i+\widehat p_j)^2$, and define
\begin{equation}
 X=\frac{(\widehat p_1^{\,2},S_{12},S_{23},S_{34},S_{45},S_{15})}{q^2}
   =(1,v_2,u_1,u_2,u_3,v_1).
 \label{eq:integral-variables}
\end{equation}
The Gram determinant is
$\Delta_5=\det(2p_i\!\cdot p_j)_{i,j=1}^4$.
The oriented root $\tr_5$ changes sign under odd permutations.
The source notation $s_{ij}$ of Ref.~\cite{integrals} is denoted here
by $S_{ij}$ to distinguish it from our form-factor invariants.
Its physical $s_{45}$ channel has incoming $\widehat p_4=p_3$ and
$\widehat p_5=p_4$, with outgoing
$(\widehat p_1,\widehat p_2,\widehat p_3)=(-q,p_1,p_2)$.
It is therefore our $s_{34}$ production channel.
In the all-outgoing convention, $p_3,p_4$ are past-directed and
$-q,p_1,p_2$ are future-directed. We choose
$\tr_5=4i\varepsilon_{\mu\nu\rho\sigma}
p_1^\mu p_2^\nu p_3^\rho p_4^\sigma$ and
$\tr_5/\sqrt{\Delta_5}=+1$ at the reference point.
The principal $\sqrt{\Delta_5}$ is permutation invariant, whereas
$\tr_5$ changes sign under odd permutations.
With $q^2=\mu^2=1$, at the reference point
\begin{equation}
 X_0=(1,3,2,-2,7,-2)
 \label{eq:source-point}
\end{equation}
we choose $\tr_5=2i\sqrt7$, with the remaining roots and boundary
constants as in Ref.~\cite{integrals}.
Thus $(u_1,u_2,u_3,v_1,v_2)=(2,-2,7,-2,3)$ at $X_0$.
Continuation from $X_0$ starts with a series expansion near that point,
using these boundary constants. These conventions specify the sheet
used for the soft, two-particle collinear, and Steinmann tests below.
The triple-collinear construction uses the decay base point
$X_{\rm dec}$ in Eq.~\eqref{eq:tc-integral-coordinates}.

\subsubsection*{Rational prefactors and numerical unitarity}
\label{sec:numerical-unitarity-sm}

Numerical generalized unitarity determines rational coefficients by
matching an integrand ansatz to products of on-shell tree objects
\cite{unitarity}. For a cut $C$, let $\ell_C$ obey the on-shell equations
$\rho_j(\ell_C)=0$ for the cut inverse propagators. The matching equation is
\begin{equation}
 \sum_{\Gamma\supset C}\sum_\alpha
 c_{\Gamma\alpha}(x,D)
 \frac{m_{\Gamma\alpha}(\ell_C)}
 {\prod_{j\in\Gamma\setminus C}\rho_j(\ell_C)}
 =\sum_{\text{states, colors}}
 \FF_{\mathcal O}^{(0)}\prod_v\mathcal A_v^{(0)}.
 \label{eq:cut-linear-system}
\end{equation}
Here $x$ denotes the external kinematics, $\Gamma$ a propagator topology,
and $m_{\Gamma\alpha}$ a chosen numerator basis. The cut removes its
propagators from every term containing $C$. Numerators proportional to a
cut propagator are invisible to that cut and are constrained on cuts
with fewer propagators. Evaluating Eq.~\eqref{eq:cut-linear-system} at
several allowed loop momenta gives a linear system for the
$c_{\Gamma\alpha}(x,D)$. Its right-hand side is computed from trees,
independently of the loop ansatz. The samples can be exact rational
numbers or finite-field elements \cite{finitefields}. They are not
numerical evaluations of loop integrals.

\paragraph*{A one-loop cut.}
The four-point one-loop calculation gives a small explicit example.
Use the scalar component with external states
$(\phi,\phi,g^+,g^+)$ and divide its cut by the tree form factor
$\langle12\rangle^2 P_1$. On the $q^2$ cut, put
$\ell_1+\ell_2=-q$ and $\ell_1^2=\ell_2^2=0$.
The name refers to the total invariant $(\ell_1+\ell_2)^2=q^2$,
not to $q^2=0$. We evaluate algebraic residues at on-shell momenta,
without performing a phase-space integral.
Figure~\ref{fig:one-loop-q2-cut} shows the ordered contribution with
the cut pair inserted between legs $4$ and $1$. With all amplitude
momenta outgoing and color tensors suppressed on the ordered trees,
the normalized sewn product is
\begin{equation}
 R(\ell_1,\ell_2)=
 \frac{\displaystyle\sum_{\mathrm{states}}
 \FF_2^{(0)}(-\ell_2,-\ell_1;q)\,
 \mathcal A_6^{(0)}(\ell_1,\ell_2,1,2,3,4)}
 {\FF_4^{(0)}(1,2,3,4;q)}.
 \label{eq:one-loop-sewn-product}
\end{equation}
The state labels on each sewn pair are conjugate under crossing.
The minimal $\tr(\phi^2)$ form factor fixes them to scalars in this
channel. Its outgoing momenta are $-\ell_1,-\ell_2$, whose sum is
$q=\sum_i p_i$, while momentum conservation on the amplitude gives
$\ell_1+\ell_2+\sum_i p_i=0$.
The corresponding tree formula is
\begin{equation}
 R(\ell_1,\ell_2)
 =-\frac{\langle\ell_1\ell_2\rangle\langle41\rangle}
 {\langle\ell_2 1\rangle\langle4\ell_1\rangle},
 \qquad
 A=(\ell_2+p_1)^2,\qquad B=(\ell_1+p_4)^2.
 \label{eq:one-loop-cut-rhs}
\end{equation}
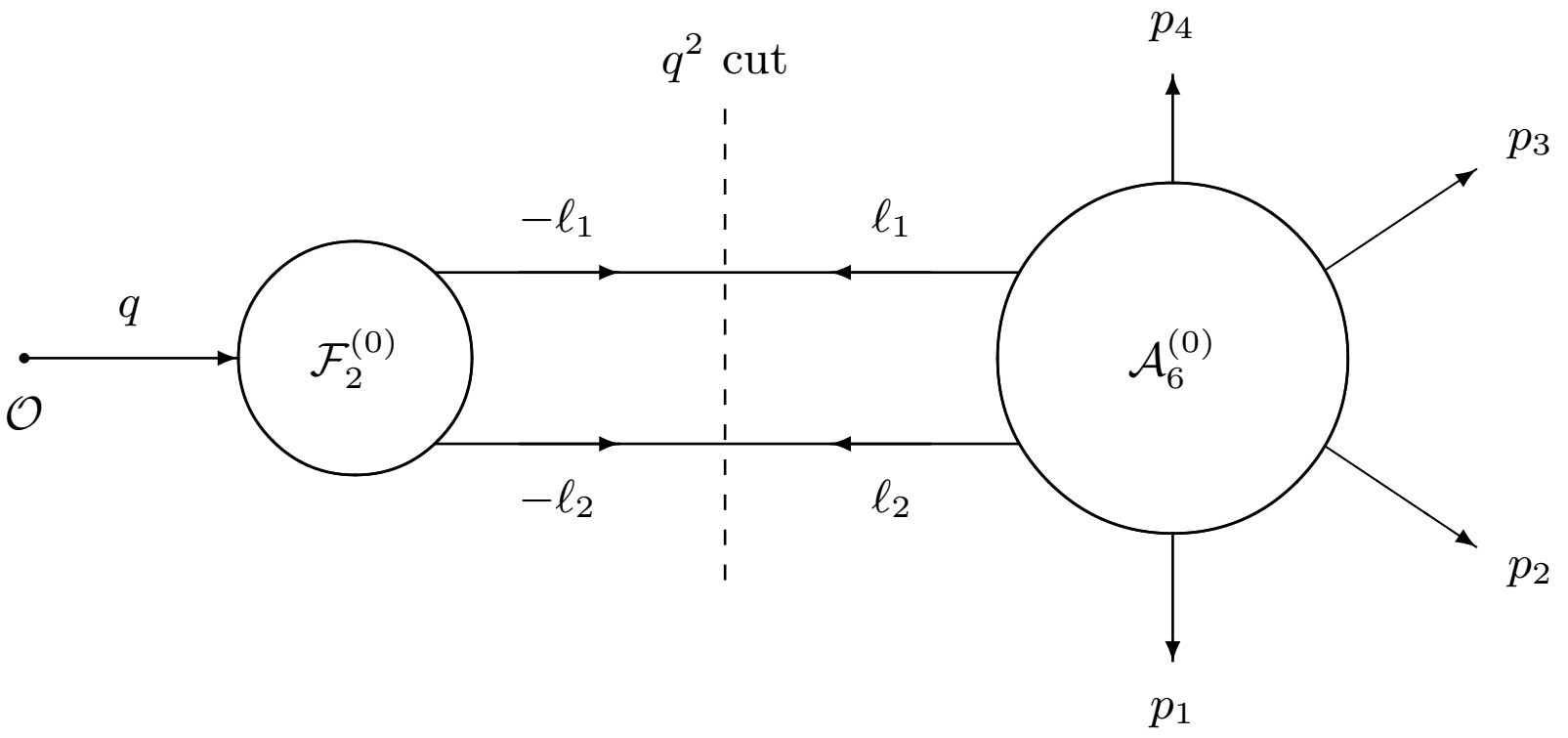
\begin{figure}[tb]
\centering
\begingroup
\setlength{\unitlength}{0.75pt}
\begin{picture}(430,190)
\linethickness{0.5pt}
\qbezier(105,130)(117.43,130)(126.21,121.21)
\qbezier(126.21,121.21)(135,112.43)(135,100)
\qbezier(135,100)(135,87.57)(126.21,78.79)
\qbezier(126.21,78.79)(117.43,70)(105,70)
\qbezier(105,70)(92.57,70)(83.79,78.79)
\qbezier(83.79,78.79)(75,87.57)(75,100)
\qbezier(75,100)(75,112.43)(83.79,121.21)
\qbezier(83.79,121.21)(92.57,130)(105,130)
\put(105,100){\makebox(0,0){$\FF_2^{(0)}$}}
\qbezier(315,145)(333.64,145)(346.82,131.82)
\qbezier(346.82,131.82)(360,118.64)(360,100)
\qbezier(360,100)(360,81.36)(346.82,68.18)
\qbezier(346.82,68.18)(333.64,55)(315,55)
\qbezier(315,55)(296.36,55)(283.18,68.18)
\qbezier(283.18,68.18)(270,81.36)(270,100)
\qbezier(270,100)(270,118.64)(283.18,131.82)
\qbezier(283.18,131.82)(296.36,145)(315,145)
\put(315,100){\makebox(0,0){$\mathcal A_6^{(0)}$}}
\put(20,100){\circle*{3}}
\put(20,100){\vector(1,0){55}}
\put(20,86){\makebox(0,0){$\mathcal O$}}
\put(47,112){\makebox(0,0){$q$}}
\put(125.4,122){\line(1,0){150.2}}
\put(125.4,78){\line(1,0){150.2}}
\put(147,122){\vector(1,0){26}}
\put(253,122){\vector(-1,0){26}}
\put(147,78){\vector(1,0){26}}
\put(253,78){\vector(-1,0){26}}
\put(157,136){\makebox(0,0){$-\ell_1$}}
\put(243,136){\makebox(0,0){$\ell_1$}}
\put(157,64){\makebox(0,0){$-\ell_2$}}
\put(243,64){\makebox(0,0){$\ell_2$}}
\multiput(200,43)(0,9){14}{\line(0,1){4}}
\put(200,177){\makebox(0,0){$q^2$ cut}}
\put(315,55){\vector(0,-1){33}}
\put(315,9){\makebox(0,0){$p_1$}}
\put(354,77.5){\vector(3,-2){39}}
\put(407,45){\makebox(0,0){$p_2$}}
\put(354,122.5){\vector(3,2){39}}
\put(407,155){\makebox(0,0){$p_3$}}
\put(315,145){\vector(0,1){28}}
\put(315,185){\makebox(0,0){$p_4$}}
\end{picture}
\endgroup
\caption{One ordered contribution to the one-loop $q^2$ cut.
The tree amplitude has cyclic order $(\ell_1,\ell_2,1,2,3,4)$.
Arrows on the sewn lines indicate outgoing momenta on each tree factor,
so $-\ell_1-\ell_2=q=\sum_i p_i$ and $\ell_1^2=\ell_2^2=0$.
The adjacent amplitude poles are $A=(\ell_2+p_1)^2$ at leg $1$ and
$B=(\ell_1+p_4)^2$ at leg $4$. The blobs denote complete tree objects.}
\label{fig:one-loop-q2-cut}
\end{figure}

The minus sign follows from the minimal form factor
$\langle\ell_1\ell_2\rangle^2/
(\langle\ell_1\ell_2\rangle\langle\ell_2\ell_1\rangle)=-1$
in this Parke--Taylor convention.
Define the parity-odd numerator
\begin{align}
 \Omega={}&\operatorname{tr}(p_1\bar q\ell_1\bar p_4)
 -2\big[(p_1\!\cdot q)(\ell_1\!\cdot p_4)
 -(p_1\!\cdot\ell_1)(q\!\cdot p_4)
 +(p_1\!\cdot p_4)(q\!\cdot\ell_1)\big],
 \label{eq:one-loop-cut-odd}
\end{align}
where the trace is over two-component momentum matrices and
$\bar p$ is the adjugate of $p$. An ansatz for this cut contribution is
\begin{equation}
 ABR=c_0+c_A A+c_B B+c_\Omega\Omega.
 \label{eq:one-loop-local-ansatz}
\end{equation}
For example, choose the rational spinors
\begin{equation}
 (\lambda_1,\lambda_2,\lambda_3,\lambda_4)
 =((1,0),(0,1),(1,2),(1,4)),\qquad
 (\widetilde\lambda_1,\widetilde\lambda_2,
 \widetilde\lambda_3,\widetilde\lambda_4)
 =((1,0),(0,1),(1,1),(2,1)).
 \label{eq:one-loop-sample-spinors}
\end{equation}
With $p_i=\lambda_i\widetilde\lambda_i^{\mathsf T}$ and
$p^2=\det p$, this point has $q^2=8$. A rational cut parametrization is
\begin{equation}
 q=\begin{pmatrix}4&2\\10&7\end{pmatrix},\qquad
 \ell_1=\begin{pmatrix}u&v\\tu&tv\end{pmatrix},\qquad
 v=\frac{-8-(7-2t)u}{4t-10},\qquad \ell_2=-q-\ell_1.
 \label{eq:one-loop-cut-parametrization}
\end{equation}
Choose $(t,u)=(0,1),(1,1),(2,1),(0,3)$ and multiply the resulting
equations by $1,6,1,5$, respectively, to clear denominators. Then
\begin{equation}
 \begin{pmatrix}
 1&-7&-8&-12\\
 6&-55&-60&-68\\
 1&-18&-20&-12\\
 5&-35&-56&-116
 \end{pmatrix}
 \begin{pmatrix}c_0\\c_A\\c_B\\c_\Omega\end{pmatrix}
 =\begin{pmatrix}56\\400\\120\\392\end{pmatrix},
 \qquad
 \begin{pmatrix}c_0\\c_A\\c_B\\c_\Omega\end{pmatrix}
 =\begin{pmatrix}2\\-2\\-7/2\\-1\end{pmatrix}.
 \label{eq:one-loop-four-by-four}
\end{equation}
Each row is the same tree identity evaluated at a different cut momentum.
The resulting generic identity is
\begin{equation}
 ABR=-\big[2(q\!\cdot p_1)(q\!\cdot p_4)
       -q^2(p_1\!\cdot p_4)\big]
       -(q\!\cdot p_4)A-(q\!\cdot p_1)B-\Omega.
 \label{eq:one-loop-cut-solution}
\end{equation}
The constant and $\Omega$ terms multiply a box denominator, while
$A$ and $B$ cancel one of its uncut propagators and therefore determine
triangle contributions. A maximal cut with $A=B=0$ cannot fix these
contact terms. The parity-odd term must be retained in the cut identity,
even though its integral for this box vanishes by Lorentz covariance.

For the complete cyclic four-dimensional example, we use eight box
topologies, with scalar and parity-odd numerators, and four triangles.
The full $q^2$-channel right-hand side sums the four cyclic placements
of the cut pair. The fitted system also averages the two assignments
$\ell_1\leftrightarrow\ell_2$ consistently on both sides.
The local example above is a single ordered, unaveraged contribution.
Five loop samples for each of the nine two-propagator cuts give
$45$ equations for these $20$ coefficients. At four external points
the system has rank $20$, and $432$ further loop samples have zero
residual. Generic rational identities also verify the three cyclic
channel types. These statements establish uniqueness within the
displayed numerator ansatz. They do not establish completeness for
arbitrary loop numerators.

\paragraph*{Dimensional terms and integral reduction.}
Four-dimensional cuts do not determine terms proportional to
$\mu_{11}=-\ell_\perp^2$. We therefore retain these terms and use
dimensionally reduced tree amplitudes. In the one-loop example, after
fixing the box and triangle terms, the $D$-dimensional $q^2$ cuts fix
four additional $\mu_{11}$ pentagons. At each of four external points,
eight samples give a rank-four system and four further samples check it,
over two prime fields. As a simple quintuple-cut example at
Eq.~\eqref{eq:one-loop-sample-spinors}, all five propagators are on shell
at $\mu_{11}=2/9$, and the sewn tree numerator gives
\begin{equation}
 \frac{2}{9}\,d_1=-\frac{8}{9},\qquad d_1=-4,
 \label{eq:one-loop-dimensional-example}
\end{equation}
where the corresponding contribution to the tree-normalized integrand is
\begin{equation}
 \frac{d_1\mu_{11}}
 {\displaystyle\prod_{r\in\{0,p_4,p_4+p_3,p_4+p_3+p_2,q\}}
 [ (\ell_{[4]}+r)^2-\mu_{11} ]},
 \label{eq:one-loop-pentagon-representative}
\end{equation}
and $\ell_{[4]}$ is the four-dimensional part of the loop momentum.
The $\mu_{11}$ numerator vanishes on every strictly four-dimensional
cut, so those cuts cannot determine $d_1$.
Independent $D$-dimensional $q^2$ tree cuts and an analytic quintuple-cut
check test this completion. The uniqueness result applies only to the
four added pentagon structures with the four-dimensional ansatz fixed,
not to arbitrary $D$-dimensional numerators.

Integration-by-parts identities then replace integrals by a chosen master
basis. This is a separate linear problem from cut matching.
For example, define
\begin{equation}
 I(a,b,c)=\int_\ell
 \frac{1}{(\ell^2)^a[(\ell+p)^2]^b[(\ell+Q)^2]^c},
 \qquad p^2=0,\quad Q^2=t,\quad(Q-p)^2=s,
 \label{eq:one-loop-triangle-family}
\end{equation}
with the measure of Eq.~\eqref{eq:coupling-measure}.
Let $T=I(1,1,1)$, $J=I(1,1,2)$,
$B_s=I(0,1,1)$, $B_t=I(1,0,1)$,
$B_s^{(2,1)}=I(0,2,1)$ and $B_t^{(2,1)}=I(2,0,1)$.
Total derivatives with vectors $\ell$ and $\ell+p$, together with
the corresponding bubble identities, give
\begin{align}
 0&=(D-4)T-2B_s^{(2,1)}+tJ,&
 0&=(D-4)T-2B_t^{(2,1)}+sJ,\\
 0&=(D-3)B_s+sB_s^{(2,1)},&
 0&=(D-3)B_t+tB_t^{(2,1)}.
 \label{eq:one-loop-ibp-system}
\end{align}
Scaleless integrals have been set to zero. Bubble symmetry identifies
which propagator is squared. Eliminating the doubled integrals yields
\begin{equation}
 T=-\frac{2(D-3)}{D-4}\frac{B_s-B_t}{s-t}
   =\frac{1-2\eps}{\eps}\frac{B_s-B_t}{s-t}.
 \label{eq:one-loop-ibp-result}
\end{equation}
Thus a triangle coefficient obtained from the cut becomes two known
bubble-master coefficients. The identity is also checked by direct
Feynman-parameter integration. Full-color coefficients follow by
crossing the ordered result and multiplying the color rings
$N S_i+2(D_1+D_2+D_3)$, so the common double-trace coefficient is
$K^{(1)}=L_1^{(1)}+L_2^{(1)}+L_3^{(1)}$. Superscripts distinguish
these one-loop coefficients from $L_i,K$ in Eq.~\eqref{eq:mainresult}.

At two loops, we apply the same separation between tree matching and
integral reduction to the coefficient combinations required by the
bootstrap. Cut-compatible integration-by-parts relations project the
sewn numerators onto master-integral directions. Surface terms can be
nonzero on a cut, despite integrating to zero, and are retained in the
cut matching. Their removal in a cut quotient is not an identity between
uncut integrands. Uncut reductions and compatible lower-cut information
are used when needed to determine the required combinations.
The construction does not require a complete symbolic integrand or an
analytic reduction of every numerator at generic external kinematics.
At each sampled point, rational dependence on $D$ is reconstructed from
twelve values and checked at four further values. The published
master-integral expansions \cite{integrals} then turn the reduced
coefficients into vectors of formal coefficients in the complete function
basis, including products.
Color-space infrared subtraction is performed on these vectors,
including all $i\pi$ and zeta terms, before the ordinary symbol is taken.
These independently obtained vectors supply the numerical equations
that fix the remaining bootstrap parameters.

\paragraph*{The third rational prefactor.}
Besides $P_1,P_2$, the regulated cuts contain
\begin{equation}
 R_3=P_1\frac{s_{12}s_{34}+s_{14}s_{23}-s_{13}s_{24}}{\tr_5}.
 \label{eq:R3}
\end{equation}
The one-loop $\mu_{11}$ pentagons contribute from order $\eps$ onward.
They cannot be dropped before two-loop subtraction, because the
one-loop infrared operator has double poles. We retain the regulated
one-loop result through $\eps^2$, including this prefactor.

We test an enlarged three-prefactor ansatz using the raw numerical-cut
and integral-reduction data, before the two-prefactor fit. A crossing
projection isolates the third coefficient in all six subleading-color
trace components at nine kinematic assignments over six prime
fields. Its nonzero bare contribution cancels coefficient by coefficient
against infrared subtraction, including products, $i\pi$, zeta values
and constants. The counterterm follows from the retained one-loop
$\eps$ and $\eps^2$ terms. An independent three-prefactor fit at three
points, checked at a fourth, gives the same vanishing coefficient.

A separate exact calculation at generic kinematics gives zero in a
rank-$90$ projection of weight-four symbols modulo products. Together,
these checks support the two-Parke--Taylor hard-function ansatz in
Eq.~\eqref{eq:mainresult}, conditional on the retained cut and reduction
space and the constant-coefficient pure-function ansatz. They do not
constitute a general prefactor-completeness theorem or a
characteristic-zero proof for the full third coefficient.

\paragraph*{Computational implementation.}
The tree currents, state sums, finite-field matching and sparse function
assembly are implemented in our own Python, C++ and Wolfram Language code.
\texttt{ColorMath}~\cite{colormath} independently verifies the general-$N$ trace
decomposition. \texttt{Singular}~\cite{singular} computes syzygies for the
cut integration-by-parts vectors, which we check by direct substitution.
\texttt{Kira}~\cite{kira} reduces the uncut integral targets, with
\texttt{Fermat}~\cite{fermat} performing rational-function arithmetic when $D$
is kept symbolic over a finite field.
\texttt{FLINT}~\cite{flint} supplies the small-prime matrix and polynomial operations.
A separate large-prime matrix interface is tested with
\texttt{FiniteFlow}~\cite{finiteflow}.
Selected reductions are independently compared with the local
\texttt{FeynmanReducer} implementation~\cite{feynmanreducer}, whose numerical interface uses
\texttt{FireFly}~\cite{firefly}.
The published master-integral expansions and their boundary constants are
imported from Ref.~\cite{integrals}.
Selected physical-point evaluations and calibrations use
\texttt{PentagonFunctions++}~\cite{pentagon-cpp} and \texttt{QD}~\cite{qd}.
The auxiliary symbolic and numerical work uses
\texttt{Mathematica}~\cite{mathematica}, \texttt{SymPy}~\cite{sympy}, \texttt{NumPy}~\cite{numpy},
\texttt{mpmath}~\cite{mpmath} and \texttt{SciPy}~\cite{scipy}.
In particular, the independent Chen-system transport uses \texttt{SciPy}'s DOP853
integrator.

GPT-6 Astra (OpenAI) assisted the alphabet and triple-collinear analyses
and the development of checking code under the authors' direction.
The resulting expressions were checked using the symbolic identities,
independent numerical evaluations and continuation tests described here.

\subsubsection*{Physical constraints}
\label{sec:physical-constraints}

Soft and two-particle collinear factorization, together with ordinary
Steinmann relations, constrain the symbol and complete-function ans\"atze.
At symbol level, matching the complete numerical-unitarity coefficient
vectors fixes the crossing-symmetric ans\"atze without imposing these
conditions. We test the reconstructed functions against them below.
We also check that the complete functions remain finite as the Gram
determinant approaches zero.
The Chen integrals used in these checks are defined in
Eq.~\eqref{eq:gpl-chen} below.

\paragraph*{Soft and collinear factorization.}

In the timelike $1\parallel2$ limit, put $s_{12}/q^2=\delta^2$.
For either a collinear or soft path, after extracting the leading
power of $\delta$, the tested components have the expansion
\begin{equation}
 H_r^B(\delta)=\sum_{k=0}^{4}c_{r,k}^{B}\log^k\delta
             +O(\delta\log^4\delta),\qquad B=\mathrm{coll},\mathrm{soft}.
 \label{eq:endpoint}
\end{equation}
The three collinear residues are
$f_{11}-f_{12}$, $f_{21}-f_{22}$ and $k_1-k_2$.
Their leading-power factorization predictions vanish.
In the soft limit the full hard function obeys
\begin{equation}
 \LP_{p_s\to0}|\HH_4^{(2)}\rangle
 =\sum_{\ell=0}^2S_{\rm hard}^{(\ell)}|\HH_3^{(2-\ell)}\rangle.
 \label{eq:full-soft-prediction}
\end{equation}
We test the four single-trace coefficients $f_{ij}$ and the six
double-trace coefficients
\begin{equation}
 2k_j+f_{1j}+f_{2j},\qquad 2k_j-f_{1j},\qquad 2k_j-f_{2j},
 \qquad j=1,2,
 \label{eq:soft-components}
\end{equation}
after resolving the two Parke--Taylor factors in each trace. Here the
functions are understood in the cyclically relabeled basis defined below.
Their nonzero right-hand sides follow from the two-loop
soft current \cite{soft}, the lower-point one-loop form factor
\cite{one-loop}, and the subtraction in Eq.~\eqref{eq:logZ}.
Finite contributions from the $\eps$-dependent normalization factors
are retained.

In the coordinates of Eq.~\eqref{eq:integral-variables}, the paths are
\begin{align}
 X_{\rm coll}(\delta)
 &=\left(1,\frac{15}{7}-\frac{\delta}{5}-\delta^2,
 \delta^2,-\frac{15}{14}-\frac{\delta}{5},
 5-\delta^2,-\frac32+\delta^2\right),\label{eq:coll-path}\\
 X_{\rm soft}(\delta)
 &=\left(1,6-5h,\frac{122\delta h}{331},-2h,6,
 \frac{122\delta h}{331}-\frac{424\delta}{331}-2h\right),
 \qquad h=\frac{1655-1774\delta}{1655-122\delta}.
 \label{eq:soft-path}
\end{align}
For each path let $P=X_B(0.02)$. We continue from $X_0$ along
$X_0+z(P-X_0)$ with the piecewise-linear path
\begin{equation}
 z:0\longrightarrow i/1000\longrightarrow1+i/1000\longrightarrow1,
 \label{eq:boundary-connector}
\end{equation}
and then follow $X_B(\delta)$ to the anchors $\delta=0.002,0.001$.
All roots are transported continuously.
Equation~\eqref{eq:soft-path} is written in the original labels, in which
$p_1\to0$. For comparison with the soft current, define
$(p'_1,p'_2,p'_3,p'_4)=(p_2,p_3,p_4,p_1)$, so that $p'_4\to0$.
This odd permutation reverses the oriented Gram root as part of the
crossing action. With $s'_{ij}=(p'_i+p'_j)^2$, the limiting invariants are
$(s'_{12},s'_{13},s'_{23})=(-2,-3,6)+O(\delta)$ and
$(s'_{14},s'_{24},s'_{34})=\delta(122,-424,-1350)/331+O(\delta^2)$.

We use the near-boundary expansion of the Chen integrals to extract
$c_{r,k}^B$ from the function values at each anchor. This determines the finite coefficient
$c_{r,0}^B$ independently of its factorization prediction.
Two integration settings,
$(\mathrm{rtol},\mathrm{atol})=(10^{-12},10^{-14})$ and
$(2.3\times10^{-14},2\times10^{-15})$, two anchor distances and local
series orders $8,12$ give eight numerical settings for each path.
Define
\begin{equation}
 E_k^B=\max_{r,s}
 \left|\widehat c_{r,k}^{B,(s)}-c_{r,k}^{B,\mathrm{fact}}\right|,
 \label{eq:boundary-residual}
\end{equation}
where $s$ labels those settings and $c^{\mathrm{fact}}$ is the
independent factorization prediction.
Table~\ref{tab:endpoint} gives the residuals, including the finite terms.

\begin{table}[htbp]
\centering
\begin{tabular}{@{}lrrrrr@{}}
\toprule
 & $k=0$ & $k=1$ & $k=2$ & $k=3$ & $k=4$\\\midrule
Collinear & $1.46\times10^{-5}$ & $4.80\times10^{-6}$ & $2.64\times10^{-7}$
 & $5.81\times10^{-10}$ & $3.68\times10^{-11}$\\
Soft & $1.93\times10^{-6}$ & $3.45\times10^{-7}$ & $1.50\times10^{-8}$
 & $1.09\times10^{-10}$ & $4.90\times10^{-12}$\\\bottomrule
\end{tabular}
\caption{Largest absolute residuals $E_k^B$ in the coefficients of
$\log^k\delta$, over the three collinear residues or ten soft coefficients
and eight numerical settings. The $k=0$ column tests the finite terms.}
\label{tab:endpoint}
\end{table}

Some collinear identities also follow analytically from the complete
Chen representations. Label the three residues by $r=1,2,K$.
For generic collinear kinematics away from other singularities,
with the square-root branches fixed by analytic continuation, we find
\begin{equation}
 c_{r,4}=c_{r,3}=c_{r,2}=0,\qquad
 d_{\parallel}c_{r,1}=0,\qquad c_{K,1}=0.
 \label{eq:exact-collinear-reduction}
\end{equation}
Here $d_{\parallel}$ differentiates with respect to the remaining
kinematic variables while keeping $p_1\parallel p_2$.
The combination $k_1-k_2$ is odd under $1\leftrightarrow2$.
At an exchange-symmetric configuration on the specified sheet,
crossing fixes $c_{K,1}=0$. Since $d_{\parallel}c_{K,1}=0$, this
coefficient stays zero throughout the same connected region of collinear
kinematics. The remaining single-logarithm constants
and finite coefficients are checked numerically as above.

\paragraph*{Steinmann relations.}

The two tested overlapping channels are
$x=S_{12}=s_{234}$ and $y=S_{15}=s_{123}$, with $q^2=1$.
We use the spectator triples $(S_{23},S_{34},S_{45})$
\begin{equation}
 \left(\frac{21}{10}+\frac{i}{5},-\frac95+\frac{3i}{20},
       \frac{71}{10}+\frac{i}{10}\right),\qquad
 \left(\frac{23}{10}+\frac{3i}{20},-\frac{17}{10}+\frac{i}{5},
       \frac{37}{5}+\frac{3i}{20}\right).
 \label{eq:steinmann-spectators}
\end{equation}
For each triple, continue linearly from $X_0$ to the point $A$ with
$x_A=16/5+i/10$ and $y_A=-19/10+i/10$.
A series expansion supplies the initial function values near $X_0$.
Next move $x$, then $y$, along the three straight segments
\begin{equation}
 z_A\longrightarrow z_A+\frac{37}{100}+2i
 \longrightarrow\rho_z+2i\longrightarrow\rho_z,
 \qquad \rho_x=\frac1{100},\quad\rho_y=\frac3{200},
 \label{eq:steinmann-connector}
\end{equation}
holding the other coordinates fixed at their current values.
Starting from this endpoint, we continue around each channel as follows:
\begin{equation}
 (x,y)=(\rho_xe^{2\pi it},\rho_y),\qquad
 (x,y)=(\rho_x,\rho_ye^{2\pi it}),\qquad 0\le t\le1.
 \label{eq:steinmann-loops}
\end{equation}
We use the same continuation path to fix the initial square-root
branches for both orders of the two loops. With these branch choices,
the region $|x|\le\rho_x$, $|y|\le\rho_y$ contains no singularities
other than $x=0$ and $y=0$.
Let $M_x$ and $M_y$ denote analytic continuation around these loops.
The resulting double discontinuity $(M_x-1)(M_y-1)g$, without division by
$(2\pi i)^2$, is smaller than $2\times10^{-5}$ for the two spectator
choices and the numerical settings tested.
Crossing transports these tests to the other ordered overlapping pairs
at permuted kinematics, with the corresponding continuation paths.

\paragraph*{Regularity at the Gram determinant.}

To establish the behavior as $\Delta_5\to0$, we examine the complete
Chen definitions of the $1291$ basis functions. Of these, $46$ contain
$W_{198}=\sqrt{\Delta_5}$ in an integration kernel. None of the $879$
functions used in our result belongs to that set.
Consider a generic configuration with $\Delta_5=0$ where all other
letters are finite and nonzero, and the arguments of the other square
roots are nonzero. Every integration differential then has a Taylor
expansion in $t=\sqrt{\Delta_5}$. Starting the integrals at a nearby
nonsingular point, with values fixed by analytic continuation, gives
Eq.~\eqref{eq:gram}. Thus the complete functions contain no logarithms
or negative powers of $\sqrt{\Delta_5}$ near this configuration.
The series may contain odd powers of $\sqrt{\Delta_5}$.

\subsection*{II. Iterated integrals, symbols, and function reconstruction}
\label{sec:function-reconstruction}

\subsubsection*{Goncharov polylogarithms and Chen integrals}
\label{sec:gpl-functions}

We define Goncharov polylogarithms by the recursion~\cite{goncharov}
\begin{equation}
 G(a_1,\ldots,a_n;z)
 =\int_0^z\frac{dt}{t-a_1}\,G(a_2,\ldots,a_n;t),
 \qquad G(;z)=1,
 \qquad G(\underbrace{0,\ldots,0}_{n};z)=\frac{\log^n z}{n!}.
 \label{eq:gpl-recursion}
\end{equation}
Paths and logarithmic branches are continued from the stated physical
region, including in endpoint limits. Products obey the shuffle
identity, with multiplicities retained when letters coincide.

The pentagon functions in Ref.~\cite{integrals} are supplied as finite
linear combinations of based Chen integrals~\cite{chen} and transcendental
constants. For a path $X(t)$ starting at $X_0$, define
\begin{align}
 I_\gamma[w_1,\ldots,w_n;t]
 &=\int_0^t dt_n\,
    \frac{d}{dt_n}\log W_{w_n}(X(t_n))
    I_\gamma[w_1,\ldots,w_{n-1};t_n],\notag\\
 I_\gamma[;t]&=1.
 \label{eq:gpl-chen}
\end{align}
Differentiation removes the last Chen letter and the first GPL index.
Individual words may depend on the path. The complete pentagon functions
satisfy integrability.
At the production point $X_0$, the letters
$W_{16},W_{17},W_{19},W_{20}$ vanish. Individual auxiliary integrals
that are singular there use the base-point regularization fixed by
the published boundary expansion of Ref.~\cite{integrals}. The
complete functions are continued from that expansion. All 56 letters
in the decay-channel construction of $G_{11}$ are nonzero at its
base point $X_{\rm dec}$.

Whenever a one-dimensional path admits a variable $t$ in which its
letters are rational functions, factor them as
\begin{equation}
 W_j(t)=C_j\prod_r(t-\alpha_{jr})^{m_{jr}},\qquad
 d\log W_j=\sum_r m_{jr}\frac{dt}{t-\alpha_{jr}}.
 \label{eq:gpl-factor-kernel}
\end{equation}
Here a negative multiplicity represents a denominator. For a path
starting at a nonsingular point $t_0$, multilinearity gives
\begin{equation}
 I[w_1,\ldots,w_n;t_0\to t]
 =\sum_{r_1,\ldots,r_n}
   \left(\prod_{j=1}^n m_{w_jr_j}\right)
   G(\alpha_{w_nr_n}-t_0,\ldots,
     \alpha_{w_1r_1}-t_0;t-t_0).
 \label{eq:gpl-chen-conversion}
\end{equation}
The GPL indices are reversed and shifted by $t_0$. The constants
$C_j$ do not enter the kernels. The pentagon boundary constants supply
the starting values.

\subsubsection*{The ordinary symbol}

The ordinary symbol is defined recursively~\cite{classical-symbols}.
For an integrable function of weight $w$ with differential
\begin{equation}
 dF^{(w)}=\sum_\alpha F_\alpha^{(w-1)}\,d\log W_\alpha,
 \qquad
 \Symb(F^{(w)})
 =\sum_\alpha\Symb(F_\alpha^{(w-1)})\otimes W_\alpha,
 \label{eq:sm-symbol-recursion}
\end{equation}
the recursion starts from
$\Symb(\log W)=W$. We write the tensor entries as letters, understood
modulo nonzero constant factors and the multiplicative identities
among their arguments. Thus $W_aW_b$ in any slot splits into the sum
of the corresponding $W_a$ and $W_b$ entries. The shuffle identity
for the functions becomes the shuffle product of their symbols.
With the convention in Eq.~\eqref{eq:gpl-chen}, the Chen word order
agrees with the symbol order. Integrability is imposed on the complete
linear combination of words.

The ordinary symbol retains the maximal sequence of logarithmic
differentials. Positive-weight constants, including $i\pi$ and zeta
values, are set to zero in this operation. Consequently, it does not
fix lower-weight functions multiplied by such constants or the
weight-four boundary constants. The reconstruction therefore proceeds
in two stages. We first fix the symbol, then determine all remaining
terms in the complete function basis of Ref.~\cite{integrals}.

\subsubsection*{Symbol spaces and numerical reconstruction}

For the direct form-factor bootstrap, the symbol ansatz is built
recursively from the physical first entries and the integrability
equations. At fixed scale, the $203$ logarithmic differentials give
$\binom{203}{2}=20503$ pairwise wedge products.
These span a $19033$-dimensional $\Q$-vector space, yielding $1470$
constant-coefficient relations.
The dimensions of the integral-generated and direct symbol spaces,
together with the total constraint counts, are reported in the Letter. Applying the same soft and ordinary Steinmann
constraints in both ansatz spaces reduces the master-integral family
from $641$ parameters to $535$ and then $267$, respectively.
The last family has $142$ parameters in the $f_{ij}$ sector and $125$
in the $K$ sector and embeds in the corresponding $431=240+191$
dimensional direct family.
The ranks are computed over $\Q$. These counts describe homogeneous
differences between solutions of the physical equations. The soft
right-hand side is nonzero, and the particular solution is checked
against it independently.

The original master-integral reconstruction matched numerical-unitarity
data after crossing and collinear factorization had reduced the ansatz
to $641=344+297$ parameters. The injectivity argument below shows that
the same complete formal symbol-coefficient data already fix all
$1767$ master-integral and $3135$ direct
parameters after crossing, within the stated two-Parke--Taylor ansatz. No collinear, soft or ordinary Steinmann equations are needed
for this uniqueness result. These data therefore also fix all
$1067=619+448$ parameters of the direct crossing and collinear family.
Imposing the additional physical constraints first gives the subfamilies
of dimensions $267$ and $431$ in Table~\ref{tab:bootstrap} and selects the
same result. When matching is performed first, the collinear, soft and
ordinary Steinmann equations provide consistency checks on the recovered
symbols.

The matching input is the complete formal symbol-coefficient vector
obtained from the infrared-subtracted numerical cut calculation and
integral reduction. Each kinematic sample therefore supplies many
coefficient equations, not merely a complex value of an integrated function.
Let $T$ denote the action of $\sigma=(12)$ on the formal symbol tensors.
This permutation exchanges $P_1,P_2$ and the color labels $1,2$, while
fixing $K$. At a sampled point $x_p$, write $p_r=P_r(x_p)$ and let
$\bm h_a$ be the symbol-coefficient vector for color combination
$a\in\{1,2,K\}$. Crossing gives
\begin{equation}
 \begin{pmatrix}\bm h_a\\T\bm h_{\sigma(a)}\end{pmatrix}
 =\begin{pmatrix}p_1&p_2\\p_2&p_1\end{pmatrix}
  \begin{pmatrix}\bm g_{a1}\\\bm g_{a2}\end{pmatrix},
 \qquad p_1^2-p_2^2\ne0.
 \label{eq:formal-inverse}
\end{equation}
Here $\bm g_{ar}=\Symb(f_{ar})$ for $a=1,2$, and
$\bm g_{Kr}=\Symb(k_r)$. In the second row,
$\sigma(1)=2$, $\sigma(2)=1$ and $\sigma(K)=K$, and
$T$ acts on the symbol tensors with the sampled $p_r$ held fixed.
Inverting this matrix separates the two prefactor coefficients.
The inversion holds on all $3135$ crossing-covariant directions of
the enlarged ansatz and is checked coefficientwise as an exact polynomial
identity after clearing $p_1^2-p_2^2$. Hence a deformation with zero
matching data must vanish. The $1767$-dimensional master-integral crossing
family is a subspace and inherits the same injectivity.
Restricting this injective map to the
$1067$-dimensional crossing and collinear family, or its
$431$-dimensional physical subfamily, cannot introduce a kernel.

Symbol coefficients are reconstructed over two primes and checked at a
third. All coordinates,
including zero coefficients, are reconstructed. In a common letter
basis, the direct form-factor bootstrap reproduces the symbols from
the master-integral bootstrap.

\paragraph*{The fixed-order planar alphabet.}
The fixed-order intersection $V(\aset_{79})\cap V(\aset_{88})$ is
spanned by the $50$ differentials whose indices remain after excluding
the set
\begin{equation}
\begin{aligned}
 R_{29}=\{&17,19,24,26,29,31,119,122,125,128,130,136,\\
          &162,164,165,167,169,171,172,173,174,175,176,178,\\
          &180,182,183,185,187\}.
\end{aligned}
\label{eq:remove29}
\end{equation}
Of these $50$ differentials, $47$ match a single planar letter, up to
inversion and a nonzero multiplicative constant in its argument.
The other three are, with
$\omega_j=d\log W_j$ and $\ell_i=d\log a_i$,
\begin{equation}
\begin{aligned}
 \omega_{134}&=-\ell_{77}+\ell_{78}+\ell_{86}-\ell_{88},\\
 \omega_{135}&=\tfrac12(\ell_{77}+\ell_{88}-\ell_{93}),\\
 \omega_{137}&=-\ell_{77}+\ell_{86}-\ell_{88}+\ell_{89}.
\end{aligned}
\label{eq:threecombinations}
\end{equation}
Thus the $47$ single-letter matches correspond to the letters in
$\aset_{79}\setminus\{W_j:j\in R_{29}\cup\{134,135,137\}\}$.
These equalities concern logarithmic differentials, which are insensitive
to multiplicative constants. The planar $88$-letter alphabet
retains the original $a_1,\ldots,a_{93}$ labels, omitting $43,44,73,76,87$.
All formulas use fixed scale and consistently chosen square-root branches.

The three orderings $1234,1243,1324$ generate the $166$-dimensional
space $V_{S_4}(\aset_{88})$.
The crossed planar MHV space $V_{S_4}(\aset_{34})$ has dimension $61$.
The quotient $V(\aset_{79})/V_{S_4}(\aset_{34})$ is spanned by the
logarithmic differentials of twelve additional rational letters
$W_{16},\ldots,W_{27}$ and six algebraic letters.

\subsubsection*{Complete functions and boundary constants}

We expand products by the shuffle identity, keeping total weight four
in each monomial and counting the weight of its constant coefficient.
We determine the remaining coefficients in Eq.~\eqref{eq:uplift} by
matching to the numerical-unitarity vectors in the complete function
basis, retaining all $i\pi$, zeta and boundary-constant terms.
The result uses $879$ pentagon functions and has $23448$ nonzero rational
coefficients. An independent prime and a new
kinematic point reproduce every coefficient in the complete function
basis, including those with explicit $i\pi$ or zeta factors. All infrared poles vanish.
Two further generalized cuts agree as rational functions of the remaining
loop variables at that point.

An alternative uniqueness test retains crossing and only a subset of
the numerical equations. The possible differences between function
completions then span $80$ directions, which the boundary equations
reduce as
\begin{equation}
80\xrightarrow{i\pi\,\mathrm{Steinmann}}71
\xrightarrow{i\pi\,\mathrm{coll.}}35
\xrightarrow{i\pi\,\mathrm{soft}}30
\xrightarrow{\mathrm{full\ coll.}}7
\xrightarrow{\mathrm{full\ soft}}1
\xrightarrow{\mathrm{full\ Steinmann}}0.
\label{eq:homogeneous}
\end{equation}
The first three steps constrain the lower-weight symbol multiplying
$i\pi$. The last three retain the full logarithms,
dilogarithms and constants. Within this ansatz, two completions satisfying
the selected numerical equations and the stated boundary equations
therefore coincide.
The reconstructed functions are separately compared with the physical
factorization predictions, including their nonzero boundary values,
in the physical-constraints discussion above.

\paragraph*{Nonclassical components.}
We apply the projection $C_{22}$ of Eq.~\eqref{eq:c22} to the symbols
of the reconstructed complete functions to compare their nonclassical
components with the crossed planar MHV and NMHV functions.
Let $(i,j;k,l)$ denote the coefficient of
$([W_i]\wedge[W_j])\wedge([W_k]\wedge[W_l])$
in $\Lambda^2(\Lambda^2V)$. Each of the following coefficients vanishes on
all $123$ planar functions in the comparison:
\begin{equation}
\begin{array}{c|rrrr}
 &f_{11}&f_{12}&f_{21}&f_{22}\\\hline
(2,9;6,18)&1&0&1&0\\
(2,10;6,18)&0&1&0&1\\
(3,7;7,16)&1&0&-2&0\\
(3,8;6,17)&0&2&0&-1
\end{array}
\qquad\det=9.
\label{eq:witness}
\end{equation}
The comparison retains all entries of the crossed planar symbols,
including those outside the $79$-letter space. The five NMHV functions are the pure coefficients of the two-loop ratio
function in Ref.~\cite{nmhv}.

\subsubsection*{Leading-color functions and comparison with the planar result}
\label{sec:leading-color-comparison}

We reconstruct the three leading-color coefficients $L_i=P_i h_i$ from
their individual numerical-unitarity and integral-reduction data, for
the orderings $(1234,1243,1324)$, with $P_3=-P_1-P_2$.
The known planar results~\cite{planar-bootstrap,planar,planar-function}
are used only for comparison after reconstruction. All connected and
factorized integral contributions are retained. Dimension-dependent numerators and the one-loop
terms through $\eps^2$ are included before infrared subtraction.
For each of the three leading-color traces, we compare the coefficients
of $\eps^{-4},\eps^{-3},\eps^{-2}$ and $\eps^{-1}$ with the color-space
infrared prediction. We use five assignments of the external kinematic
variables, each in a different finite field $\mathbb F_p$, where rational
arithmetic is performed modulo the prime $p$. These are the four
reconstruction samples and the independent validation sample specified
below. Transcendental functions and periods remain formal basis elements,
and their rational coefficients are compared separately.
All $4\times3\times5=60$ pole differences vanish in the corresponding
finite fields.

Let $M_i^{(\ell)}$ denote the coefficient of $N^\ell S_i$ in
$\FF^{(\ell)}$, divided by $P_i$, with the common MHV numerator
suppressed. For ordering $i$, define
$A_i=\sum_{(a,b)\,\mathrm{adjacent}}\log[(-s_{ab}-i0)/\mu^2]$,
where the sum contains the four cyclic adjacent pairs. Write
\begin{align}
 M_i^{(1)}={}&-\frac4{\eps^2}+\frac{A_i}{\eps}
       +h_i^{(1)}+\eps b_{i,1}+\eps^2b_{i,2}+O(\eps^3),\notag\\
 h_i={}&[M_i^{(2)}]_{\eps^0}+4b_{i,2}-A_i b_{i,1}.
 \label{eq:leading-finite-transfer}
\end{align}
The second line follows from the minimal-pole subtraction in
Eq.~\eqref{eq:hard}. The BDS convention of Ref.~\cite{planar} uses
$f^{(2)}(\eps)=-2\zeta_2-2\eps\zeta_3-2\eps^2\zeta_4$
and no additional $C^{(2)}$ subtraction. Thus
\begin{align}
 \mathcal R_{i,\mathrm{BDS}}^{(2)}
 &\equiv\lim_{\eps\to0}\left[
 M_i^{(2)}-\tfrac12(M_i^{(1)})^2
 -f^{(2)}(\eps)M_i^{(1)}(2\eps)\right]\notag\\
 &=h_i-\tfrac12\bigl(h_i^{(1)}\bigr)^2
       +2\zeta_2h_i^{(1)}+\zeta_3A_i-2\zeta_4.
 \label{eq:leading-bds-conversion}
\end{align}
This conversion fixes both the normalization and the additive constant.

After division by $P_i$, we assume constant rational coefficients in
the complete weight-four function basis. The numerical input supplies
formal coefficient vectors, without evaluating the transcendental
functions in finite fields. Reconstruction uses the primes
$1000003$, $1000039$, $250000019$ and $250001027$, whose product is $M$.
The prescribed denominator $D_c=831825068414199552000$ comes from the
subleading-color reconstruction. We require each coefficient $c$ to
have denominator dividing $D_c$ and $|D_c c|\le(M-1)/2$.
At a different kinematic point, the additional prime $1000099$, which
is not used in the reconstruction, reproduces every coefficient.
Uniqueness is within these prefactor,
denominator and height assumptions. Crossing, collinear and
triple-collinear equations are not used in this fit.

At $q^2=\mu^2=1$, exact rational arithmetic verifies
\begin{align}
 h_1-h_3&=k_1,& h_2-h_3&=k_2,&
 \sum_iL_i&=P_1k_1+P_2k_2=K,\notag\\
 h_2&=(34)h_1,& h_3&=(23)h_1.
 \label{eq:leading-K-check}
\end{align}
The crossings use the universal master-function maps with consistent
branches and also hold for the BDS remainders. Finally, all $4236$
monomial coefficients of $\mathcal R_{1,\mathrm{BDS}}^{(2)}$ agree
exactly over $\Q$ with \texttt{4\_Remainder\_Function\_in\_F.m},
the ancillary expression of Ref.~\cite{planar} (version~3).
Crossing establishes the same complete-function equality for the other
two orderings. Every $i\pi$, zeta and boundary-constant term is retained.
The first remainder's symbol also matches all $5828$ words of the
symbol of that expression, with the other two matches following by crossing.
The ancillary data supply the nine pure functions
$f_{11},f_{12},f_{21},f_{22},k_1,k_2,h_1,h_2,h_3$ and their symbols,
with the Parke--Taylor factors and color tensors specified separately.
Every symbol uses the common head \texttt{S[W[i],W[j],W[k],W[l]]}.
The files use the minimal-pole hard-function convention of
Eq.~\eqref{eq:hard}.

\subsubsection*{Complete-function check in the decay channel}
\label{sec:generic-decay-benchmark}

We evaluate all six complete coefficients in the physical decay channel
$q\to p_1+p_2+p_3+p_4$, with future-directed final-state momenta and
$q^2=\mu^2=1$. We choose the point $X_{\mathrm D}$ with
\begin{equation}
 (u_1,u_2,u_3,v_1,v_2)
 =\left(\frac{75}{484},\frac{1169}{7260},\frac{463}{2420},
         \frac{313}{660},\frac{23}{44}\right).
 \label{eq:generic-decay-check-point}
\end{equation}
In the source coordinates $X=(1,v_2,u_1,u_2,u_3,v_1)$, this is
$X_{\mathrm D}=(1,23/44,75/484,1169/7260,463/2420,313/660)$.
All six pair invariants are positive. The Gram determinant and oriented
root are
\begin{equation}
 \Delta_5=-\frac{647948019}{283449760000},\qquad
 \tr_5=\frac{i\sqrt{647948019}}{532400},\qquad
 \chi=\frac{\tr_5}{\sqrt{\Delta_5}}=+1.
 \label{eq:generic-decay-check-root}
\end{equation}
The conversion to the decay basis retains all explicit $i\pi$ and zeta
terms~\cite{decay-functions,decay-data}.
We use the Feynman prescription $\log(-s-i0)=\log s-i\pi$ for $s>0$.
The Chen solution is initialized by the boundary expansion at the
published decay reference point
$X_{\rm dec}=(1,1/2,1/6,1/6,1/6,1/2)$ and transported along the
straight real segment to $X_{\mathrm D}$. The native evaluator follows
the same segment.

We compare two implementations of the complete-function evaluation.
The native decay evaluator in
\texttt{PentagonFunctions++}~\cite{decay-functions,pentagon-cpp}
uses one-fold integrals with double-double and quad-double arithmetic
from \texttt{QD}~\cite{qd}. We contract its function values with the
reconstructed coefficient polynomials at 100-digit working precision
in \texttt{mpmath}~\cite{mpmath}.
The second implementation integrates the Chen differential system with
DOP853 in \texttt{SciPy}~\cite{scipy}, using binary64 arithmetic for the
ODE and extended precision for the coefficient evaluation and final
contraction.
The largest absolute difference between the native quad-double and
Chen results is below $1.08\times10^{-11}$ across the six coefficients.
Increasing the native evaluator's precision from double-double to
quad-double changes the contracted coefficients by less than
$1.89\times10^{-25}$, while refining the Chen integration changes them
by less than $5.09\times10^{-12}$.
These numerical comparisons support the eight-decimal values in
Table~\ref{tab:generic-decay-benchmark}.

\begin{table}[htbp]
\centering
\setlength{\tabcolsep}{5pt}
\begin{tabular}{lrrrrrr}
\hline\hline
 & $f_{11}$ & $f_{12}$ & $f_{21}$ & $f_{22}$ & $k_1$ & $k_2$ \\
\hline
$\operatorname{Re}$ & $-20.52910397$ & $2.05876713$ & $1.80126418$
 & $-21.05022743$ & $9.04625702$ & $11.03146307$ \\
$\operatorname{Im}$ & $-64.24678238$ & $-127.82217441$ & $128.63419947$
 & $64.93618745$ & $-8.32716542$ & $-10.05892651$ \\
\hline\hline
\end{tabular}
\caption{Complete coefficient functions at the decay point
$X_{\mathrm D}$ in Eq.~\eqref{eq:generic-decay-check-point}, with
$q^2=\mu^2=1$ and the root orientation in
Eq.~\eqref{eq:generic-decay-check-root}. All terms beyond the symbol are included.}
\label{tab:generic-decay-benchmark}
\end{table}

The two implementations share the published production-to-decay
conversion and decay boundary constants~\cite{decay-data}, as well as
the reconstructed coefficient expressions. Their agreement tests the
numerical evaluation, transport and assembly of the six complete
coefficients on this decay sheet. It is not an independent computation
of the form factor or a proof of the conversion map. The precision
comparisons are stability checks rather than rigorous error bounds.


\subsection*{III. Triple-collinear limit and complete $N^0$ splitting functions}
\label{sec:triple-collinear}

\subsubsection*{Kinematics, continuation, and hard factorization}

We take $p_1,p_2,p_3$ collinear at fixed nonzero $q^2$ and introduce
\begin{equation}
 \rho=\frac{s_{123}}{q^2},\qquad
 x=\frac{s_{12}}{s_{123}},\qquad y=\frac{s_{23}}{s_{123}},\qquad
 a=z_1,\qquad b=z_3,\qquad c=1-a-b,\qquad h=1-x-y.
 \label{eq:tc-kinematics}
\end{equation}
At fixed shape $\xi=(x,y,a,b)$, we approach the limit along
\begin{equation}
 \frac{s_{12}}{q^2}=\rho x,\qquad
 \frac{s_{23}}{q^2}=\rho y,\qquad
 \frac{s_{13}}{q^2}=\rho h,\qquad
 \frac{s_{i4}}{q^2}=(1-\rho)z_i,
 \qquad (z_1,z_2,z_3)=(a,c,b).
 \label{eq:tc-invariant-path}
\end{equation}
We set $q^2=\mu^2=1$ and define the oriented shape root by
$d=\tr_5/[\rho(1-\rho)]$.
In the physical decay region all fractions in
Eq.~\eqref{eq:tc-kinematics} are positive and
\begin{equation}
 d^2=\lambda(bx,ch,ay)<0,\qquad
 \lambda(A,B,C)=A^2+B^2+C^2-2AB-2AC-2BC.
 \label{eq:tc-gram-shape}
\end{equation}
We use the decay continuation of the complete form factor, with
positive imaginary $d$ on the reference orientation
$\tr_5/\sqrt{\Delta_5}=+1$.

We first apply the exact
production-to-decay map of Refs.~\cite{decay-functions,decay-data} to
the complete expressions. The map acts on functions, including their
$i\pi$ terms and products. The weight-one scale functions
$F^{\rm prod}_{1,1}$ and $F^{\rm dec}_{1,1}$ in these bases obey
\begin{equation}
 F^{\rm prod}_{1,1}=F^{\rm dec}_{1,1}+i\pi.
 \label{eq:gpl-scale-map}
\end{equation}
At $q^2=1$ the production function on the left vanishes, while
$F^{\rm dec}_{1,1}=-i\pi$. We substitute this decay value in the
continued polynomials, retaining its contributions to every weight.
The decay base point is specified in
Eq.~\eqref{eq:tc-integral-coordinates}. All square-root branches are
continued from that point. Odd external permutations also change the oriented Gram
invariant when applying Eq.~\eqref{eq:seed}.

For the minimal-pole hard functions of Eq.~\eqref{eq:hard}, timelike
factorization gives
\begin{equation}
 \LP\HH_4^{(2)}
 =\mathrm{Sp}_H^{(2)}\HH_2^{(0)}
  +\mathrm{Sp}_H^{(1)}\HH_2^{(1)}
  +\mathrm{Sp}^{(0)}\HH_2^{(2)},\qquad
 \mathrm{Sp}_H=\lim_{\epsilon\to0}\mathbf Z_4^{-1}
                 \mathrm{Sp}\,\mathbf Z_2.
 \label{eq:tc-hard-factorization}
\end{equation}
We write the coupling $a$ of Eq.~\eqref{eq:coupling-measure} as
$a_{\rm loop}$ here. The two-point one- and two-loop functions carry
$N\delta^{a_Pa_4}$ and $N^2\delta^{a_Pa_4}$, respectively. The tree
splitting has single-trace color, and the one-loop splitting has
$N$ times single traces and unweighted double traces. Consequently, the
last two terms of Eq.~\eqref{eq:tc-hard-factorization} contribute only
to $N^2S$ and $ND$. The $N^0$ coefficients $f_{ij}$ therefore need no
lower-point Sudakov subtraction, whereas the $ND$ coefficients do.

The four-point normalization fixes
$\mathrm{Sp}_H^{(2)}\HH_2^{(0)}$. Extracting a splitting current
requires the two-point tree normalization and the Parke--Taylor factors.
Writing $P=p_1+p_2+p_3$ at leading power and
$P_{P4}=1/(\langle P4\rangle\langle4P\rangle)$, one obtains
\begin{equation}
 \frac{P_1}{P_{P4}}\longrightarrow
 T_1=\frac{1}{\sqrt{ab}\,\langle12\rangle\langle23\rangle},\qquad
 \frac{P_2}{P_{P4}}\longrightarrow
 T_2=\frac{1}{\sqrt{bc}\,\langle12\rangle\langle31\rangle},\qquad
 T_3=-T_1-T_2.
 \label{eq:tc-pt-limits}
\end{equation}
Thus the coefficient of $S_1$ obeys
\begin{equation}
 \frac{C_1}{P_1}\longrightarrow f_{11}+R f_{12},\qquad
 R=\frac{T_2}{T_1}
  =-\sqrt{\frac ac}\frac{\langle23\rangle}{\langle13\rangle},\qquad
 chR^2+(ch+ay-bx)R+ay=0.
 \label{eq:tc-physical-trace}
\end{equation}
The two roots correspond to the oriented Gram branches and must be
transported with the spinors. A chosen component also requires its
MHV numerator. For scalar legs $1,4$, its ratio to the two-point
numerator is $a$. Thus $f_{11}$ is one pure coefficient entering this contracted
splitting contribution.

\subsubsection*{Complete scale logarithms}

Define
\begin{equation}
 L=\log\frac{-s_{123}-i0}{\mu^2}=\log\rho-i\pi,\qquad
 U=\log\frac{ay}{ch},\qquad V=\log\frac{bx}{ch}.
 \label{eq:tc-log-definitions}
\end{equation}
The complete leading-power expansions of the four $N^0$ pure functions
are
\begin{align}
 f_{11}^{\rm LP}&=V(U+V)L^2-2(U+V)\kappa_1L+G_{11},\notag\\
 f_{12}^{\rm LP}&=(V-U)(U+V)L^2-2(U+V)\kappa_2L+G_{12},\notag\\
 f_{21}^{\rm LP}&=V(V-2U)L^2-2(V-2U)\kappa_1L+G_{21},\notag\\
 f_{22}^{\rm LP}&=(V-U)(V-2U)L^2-2(V-2U)\kappa_2L+G_{22}.
 \label{eq:tc-full-logarithms}
\end{align}
The difference from the full functions is $o(1)$ at fixed generic
shape. We write the weight-four finite terms as $G_{ij}(x,y,a,b)$,
with $a=z_1$ and $b=z_3$ in the notation of the Letter. Define
\begin{equation}
 \kappa_1=\beta(x,y;a,b)-\beta(h,y;a,c),\qquad
 \kappa_2=\beta(h,x;b,c)-\beta(h,y;a,c),
 \label{eq:tc-kappa}
\end{equation}
where $z_2=1-z_1-z_3$ and
\begin{align}
 \beta(X,Y;z_1,z_3)={}&-\frac12\log^2(XY)
 +\log\frac{1-z_3}{z_1}\log X
 +\log\frac{1-z_1}{z_3}\log Y\notag\\
 &+\operatorname{Li}_2\!\left(-\frac{z_2}{z_1}\right)
  +\operatorname{Li}_2\!\left(-\frac{z_2}{z_3}\right)
  +\operatorname{Li}_2\!\left(-\frac{z_3}{1-z_3}\right)
  +\operatorname{Li}_2\!\left(-\frac{z_1}{1-z_1}\right)\notag\\
 &-\operatorname{Li}_2\!\left(1-\frac{X}{1-z_3}\right)
  -\operatorname{Li}_2\!\left(1-\frac{Y}{1-z_1}\right).
 \label{eq:tc-beta}
\end{align}
This expression follows from the one-loop MHV triple splitting
amplitude~\cite{splitting-one-loop}. A common additive weight-two
constant cancels from Eq.~\eqref{eq:tc-kappa}. The functions displayed
in Eqs.~\eqref{eq:tc-kappa} and \eqref{eq:tc-beta} are real in the
physical interior. The scale phase is carried by the prescribed $L$.

The absence of $L^4$ and $L^3$ follows at function level from color
evolution. Let $\mathbf T_{ij}=\mathbf T_i\cdot\mathbf T_j$.
Differentiating Eq.~\eqref{eq:logZ} gives the $n$-adjoint anomalous
dimension through two loops,
\begin{equation}
 \boldsymbol\Gamma_n
 =2a_{\rm loop}\mathbf M_n
  +a_{\rm loop}^2\bigl(-4N\zeta_2\mathbf M_n
                         +2nN^2\zeta_3\mathbf1\bigr),\qquad
 \mathbf M_n=-2\sum_{i<j}\mathbf T_{ij}
                    \log\frac{-s_{ij}-i0}{\mu^2}.
 \label{eq:tc-anomalous-dimension}
\end{equation}
Color conservation implies
$\sum_{1\leq i<j\leq3}\mathbf T_{ij}=-N$ on the parent-adjoint
channel. Subtracting the two-point parent-spectator contribution
leaves
\begin{equation}
 \mathbf M_{\rm sp}=2NL+\mathbf M_0,\qquad
 \mathbf M_0=-2\left[
  \mathbf T_{12}\log(bx)+\mathbf T_{13}\log(ch)
  +\mathbf T_{14}\log(ay)\right].
 \label{eq:tc-splitting-color-log}
\end{equation}
Opposite-channel dipole matrices are equal on the color-singlet
four-leg trace space, which permits the form of
Eq.~\eqref{eq:tc-splitting-color-log}.

After removing the overall collinear power, let $v$ be the tree color
vector and $S_H^{(\ell)}$ represent
$\mathrm{Sp}_H^{(\ell)}\HH_2^{(0)}$. Then
\begin{align}
 \partial_LS_H^{(1)}&=-(2NL+\mathbf M_0)v,\notag\\
 \partial_LS_H^{(2)}&=-(2NL+\mathbf M_0)S_H^{(1)}
 +\left[2N\zeta_2(2NL+\mathbf M_0)-2N^2\zeta_3\right]v.
 \label{eq:tc-scale-equations}
\end{align}
Integration gives $S_H^{(1)}=-NL^2v-L\mathbf M_0v+B_1$.
Since $v$ has color $S$ and $B_1$ has colors $NS$ and $D$, the two-loop
projection is
\begin{equation}
 [S_H^{(2)}]_{N^0S}
 =\tfrac12L^2[\mathbf M_0^2v]_{N^0S}
  -L[\mathbf M_0B_1]_{N^0S}+[B_2]_{N^0S}.
 \label{eq:tc-nzero-projection}
\end{equation}
Here $B_1,B_2$ are scale independent. The double-trace part of $B_1$
enters the linear term. The two-loop cusp and diagonal terms cannot
contribute to $N^0S$, excluding cubic or quartic logarithms and
corrections to the quadratic coefficient that are invisible to the symbol.

The required color algebra is short. Put
$\mathfrak a=\log(bx)$, $\mathfrak b=\log(ch)$,
$\mathfrak c=\log(ay)$, and use the ordered basis
$(S_1,S_2,S_3,D_1,D_2,D_3)$. Then
\begin{equation}
 (\mathbf M_0)_{S\leftarrow D}
 =\begin{pmatrix}
 \mathfrak c-\mathfrak b&0&\mathfrak a-\mathfrak b\\
 \mathfrak b-\mathfrak c&\mathfrak a-\mathfrak c&0\\
 0&\mathfrak c-\mathfrak a&\mathfrak b-\mathfrak a
 \end{pmatrix}.
 \label{eq:tc-color-block}
\end{equation}
For $v=P_1S_1+P_2S_2+P_3S_3$ and $P_3=-P_1-P_2$, the double-trace
entries of $\mathbf M_0v$ all equal $-2Q$, where
$Q=(\mathfrak b-\mathfrak a)P_1+
(\mathfrak c-\mathfrak a)P_2$. Hence
\begin{equation}
 \tfrac12[\mathbf M_0^2v]_{N^0S}
 =-Q\begin{pmatrix}
 \mathfrak a-2\mathfrak b+\mathfrak c\\
 \mathfrak a+\mathfrak b-2\mathfrak c\\
 -2\mathfrak a+\mathfrak b+\mathfrak c
 \end{pmatrix},\qquad
 (B_1)_{D_i}=2(P_1\kappa_1+P_2\kappa_2)\quad(i=1,2,3).
 \label{eq:tc-color-evaluation}
\end{equation}
Substitution into Eq.~\eqref{eq:tc-nzero-projection} gives the scale
logarithms in Eq.~\eqref{eq:tc-full-logarithms}, including their terms
beyond the symbol.
The ordinary symbols agree with the direct triple-collinear limit.

Writing $\ell=\log\rho$, a function
$f^{\rm LP}=R_2L^2+R_1L+G$ obeys
\begin{equation}
 f^{\rm LP}=R_2\ell^2+(R_1-2i\pi R_2)\ell+C_0,
 \qquad G=C_0+i\pi R_1+\pi^2R_2.
 \label{eq:tc-finite-part-phase}
\end{equation}

\subsubsection*{Finite alphabet}

We call alphabet letters independent when their logarithmic differentials
are linearly independent over $\mathbb Q$, disregarding multiplicative
constants in the letters. The alphabets below include the
lower-weight terms multiplying transcendental constants in $G_{ij}$.

\paragraph*{The 29 finite letters.}
Introduce the triples
\begin{equation}
 r=(r_1,r_2,r_3)=(y,h,x),\qquad
 z=(z_1,z_2,z_3)=(a,c,b),\qquad
 \sum_{i=1}^{3}r_i=\sum_{i=1}^{3}z_i=1.
 \label{eq:tc-alph-rz}
\end{equation}
The rational block is
\begin{equation}
\begin{aligned}
 \mathcal A_{\mathrm{rat}}
 &=\{z_i,1-z_i,r_i,1-r_i,1-z_i-r_i:\ i=1,2,3\}\\
 &=\{a,b,c,1-a,1-b,a+b,x,y,h,1-x,1-y,x+y,\\
 &\hspace{2.4em}1-a-y,1-b-x,a+b+x+y-1\}.
\end{aligned}
 \label{eq:tc-alph-rational}
\end{equation}
These 15 factors are distinct and irreducible. Besides the Gram root
$d$, the alphabet contains $\eta_1,\eta_4,\eta_6$, denoted
$h_1,h_4,h_6$ in the ancillary maps. Define
\begin{equation}
\begin{aligned}
 A_1&=(1-a)(1-x)-b(1-y),\\
 A_4&=a(1-x)-b(1-y),\qquad B_4=a(1-x)+b(1-y),\\
 A_6&=a(1-x)-(1-b)(1-y),
\end{aligned}
 \label{eq:tc-alph-linear-root-parts}
\end{equation}
and set
\begin{equation}
\begin{aligned}
 d^2&=\lambda(xb,hc,ya),\\
 \eta_1^2&=A_1^2+4(1-a)bxy,\\
 \eta_4^2&=B_4^2-4abxy,\\
 \eta_6^2&=A_6^2+4a(1-b)xy.
\end{aligned}
 \label{eq:tc-alph-roots}
\end{equation}
Here $\lambda$ is defined in Eq.~\eqref{eq:tc-gram-shape}, and
$\sqrt{\Delta_5}=\rho(1-\rho)d$ on the reference orientation.
We continue from positive imaginary $d$ and positive real $\eta_i$
in the decay region. Changing a root sign inverts its algebraic ratios.

For the Gram block define
\begin{equation}
 C=(1-a)(1-b),\qquad
 u=\frac{x}{1-b},\qquad v=\frac{y}{1-a},\qquad
 w=\frac{ab}{C},\qquad
 \Delta=(1-u-v-w)^2-4uvw.
 \label{eq:tc-alph-cross-ratios}
\end{equation}
Then $d^2=C^2\Delta$.  With the ratio
$\Phi(A,R)=(A+R)/(A-R)$, choose
\begin{equation}
\begin{aligned}
 y_u&=\frac{C(1+u-v-w)-d}{C(1+u-v-w)+d},&
 y_v&=\frac{C(1-u+v-w)-d}{C(1-u+v-w)+d},\\
 y_w&=\frac{C(1-u-v+w)-d}{C(1-u-v+w)+d},&
 E_1&=\Phi(xb-ya-hc,d),\\
 &&E_2&=\Phi(hc-ya-xb-2ax,d),
\end{aligned}
 \label{eq:tc-alph-gram-letters}
\end{equation}
so that $\mathcal A_{\mathrm G}=\{y_u,y_v,y_w,E_1,E_2\}$ has five
independent letters. The three remaining blocks are explicitly
\begin{equation}
\begin{aligned}
 \mathcal A_{\Sigma_1}=
 \bigl\{&\Phi(A_1,\eta_1)^{-1},
          \Phi(A_1-2(1-a)y,\eta_1),
          \Phi(A_1-2by,\eta_1)\bigr\},\\
 \mathcal A_{\Sigma_4}=
 \bigl\{&\Phi(-B_4+2bx,\eta_4),
          \Phi(-B_4+2ay,\eta_4),
          \Phi(A_4,\eta_4)\bigr\},\\
 \mathcal A_{\Sigma_6}=
 \bigl\{&\Phi(-A_6-2(1-b)x,\eta_6),
          \Phi(A_6,\eta_6),
          \Phi(A_6+2ax,\eta_6)\bigr\}.
\end{aligned}
 \label{eq:tc-alph-sigma-letters}
\end{equation}
In the order displayed, these are the leading letters with published
indices $(162,164,183)$, $(168,170,181)$, and $(174,176,180)$, respectively.
The Gram choices satisfy, modulo constant factors,
\begin{equation}
\begin{aligned}
 y_u&=\mathcal L_{131},\qquad
 y_v=\frac{\mathcal L_{132}}{\mathcal L_{137}},\qquad
 y_w=\frac{\mathcal L_{137}}
 {\mathcal L_{130}\mathcal L_{131}\mathcal L_{132}},\\
 E_1&=\mathcal L_{132},\qquad E_2=\mathcal L_{136}.
\end{aligned}
 \label{eq:tc-alph-gram-map}
\end{equation}
Here $\mathcal L_j$ denotes the finite leading factor of the original
letter $W_j$.

The common finite alphabet is therefore the explicit union
\begin{equation}
 \mathcal A_{\mathrm{tc}}
 =\mathcal A_{\mathrm{rat}}\cup\mathcal A_{\mathrm G}
  \cup\mathcal A_{\Sigma_1}\cup\mathcal A_{\Sigma_4}
  \cup\mathcal A_{\Sigma_6},\qquad
 \dim_{\mathbb Q}V(\mathcal A_{\mathrm{tc}})=15+5+3+3+3=29.
 \label{eq:tc-alph-common}
\end{equation}
The four radicands are pairwise coprime irreducible polynomials, so their
square classes are independent over $\mathbb Q(a,b,x,y)$.
Independent root sign changes separate the algebraic blocks from
one another and from the rational block. Exact evaluations of the
logarithmic derivatives have ranks $5,3,3,3$. Together with the
15 rational divisors, these prove
Eq.~\eqref{eq:tc-alph-common}.

\paragraph*{Individual alphabets and the complete functions.}
The alphabets of the finite pure coefficients are
\begin{equation}
\begin{aligned}
 \mathcal A_{11}=\mathcal A_{21}
 &=\mathcal A_{\mathrm{rat}}\cup\mathcal A_{\mathrm G}
   \cup\mathcal A_{\Sigma_4}\cup\mathcal A_{\Sigma_6},\\
 \mathcal A_{12}=\mathcal A_{22}
 &=\mathcal A_{\mathrm{rat}}\cup\mathcal A_{\mathrm G}
   \cup\mathcal A_{\Sigma_1}\cup\mathcal A_{\Sigma_6}.
\end{aligned}
 \label{eq:tc-alph-individual}
\end{equation}
To establish the upper bounds, we expand each complete $f_{ij}$ in
decay-channel Chen integrals, retaining all source constants and
explicit $i\pi$ and zeta terms. Each uses 56 original letters, with
64 in their union. Their leading forms are
\begin{equation}
 W_\alpha(\rho,\xi)
 =\rho^{\nu_\alpha}c_\alpha\mathcal L_\alpha(\xi)
  [1+O(\rho)],\qquad \xi=(x,y,a,b).
 \label{eq:tc-alph-leading-map}
\end{equation}
Here $c_\alpha$ are nonzero constants.
For each function the leading logarithmic differentials have rank 27
including $\rho$, and rank 26 on shape space, with basis
Eq.~\eqref{eq:tc-alph-individual}. This holds separately for every
source-constant monomial, without relations among the constants.

We decompose the path into a radial segment at fixed regular shape
$\xi_*$, on which only $\rho=s_{123}/q^2$ varies, followed by a shape
segment at fixed $\rho$. Chen concatenation splits each word into a
prefix along the radial segment and a suffix along the shape segment.
The kernels along the first segment have the form
$\nu_\alpha d\log\rho$ plus regular terms, so the prefixes are
polynomials in $\log\rho$ at leading power. On a compact regular shape
path, the suffix kernels approach $d_\xi\log\mathcal L_\alpha$, and
their $O(\rho)$ errors vanish even when multiplied by powers of
$\log\rho$. Thus all logarithmic coefficients, including $G_{ij}$,
use the alphabets in Eq.~\eqref{eq:tc-alph-individual}. Radial constants
are retained even for letters with constant leading factor. The phase
correction in Eq.~\eqref{eq:tc-finite-part-phase} stays in the same
alphabet. This argument uses path independence only for the complete
function. Concatenation is valid for individual path-dependent words.

For the lower bounds, all $\rho$-dependent words cancel from the
$\Ctwotwo$ projections. Their minimal letter spaces have
dimensions 26 individually and 29 jointly, as established by exact
rational annihilators and nonzero minors.
Any smaller full-function alphabet would place these projected symbols
in tensor powers of a smaller space. The bounds therefore coincide,
proving minimality for the pure coefficients in
Eq.~\eqref{eq:tc-alph-individual}. A Parke--Taylor-weighted color
combination requires a separate analysis.

\subsubsection*{Finite functions and crossing}

The decay data supply logarithms and dilogarithms at weights one and
two and based integrals at weights three and four, with their source
constants. For the finite function $G_{11}$, the four-variable dependence is expressed
by Chen integrals over algebraic letters. Its boundary construction
uses one-dimensional integrals in $\rho$ at fixed shape. These integrals
reduce to GPLs by Eq.~\eqref{eq:gpl-chen-conversion}, with the changes
of variable given below. A global rationalization of the four-variable
alphabet is not required.

Expressing $f_{11}$ in the decay-channel pentagon
basis~\cite{decay-functions,decay-data} gives
\begin{equation}
 f_{11}(X)=\sum_{w,\gamma}q_{w,\gamma}\,\gamma\,
 I_w(X_{\rm dec}\longrightarrow X),\qquad q_{w,\gamma}\in\mathbb Q.
 \label{eq:tc-full-chen-sum}
\end{equation}
Here $\gamma$ is a monomial in the source constants, $|w|\leq4$, and
the empty word is included. We use Eq.~\eqref{eq:gpl-chen}, with
$I_w(X_{\rm dec}\to X_{\rm dec})=0$ for nonempty $w$.

In the variables of Eq.~\eqref{eq:integral-variables}, the source and
collinear path are
\begin{align}
 X_{\rm dec}&=(1,\tfrac12,\tfrac16,\tfrac16,\tfrac16,\tfrac12),\notag\\
 X(\rho,\xi)&=
 \bigl(1,\rho y+(1-\rho)(1-a),\rho x,\rho y,(1-\rho)b,\rho\bigr).
 \label{eq:tc-integral-coordinates}
\end{align}
This decay source differs from the production source
in Eq.~\eqref{eq:source-point} and satisfies $X_{\rm dec}=X(1/2,\xi_*)$,
where
\begin{equation}
 \xi_*=(\tfrac13,\tfrac13,\tfrac13,\tfrac13).
 \label{eq:tc-symmetric-shape}
\end{equation}
For the boundary constants, we vary $\rho$ from $1/2$ toward $0^+$
along $X(\rho,\xi_*)$, keeping $x=y=a=b=1/3$. This path starts at the
decay source itself. The endpoint constants are the finite parts after
the powers of $\log\rho$ have been separated.

Using the leading factors $\mathcal L_j$ in
Eq.~\eqref{eq:tc-alph-leading-map}, fix a nonsingular shape path $\eta$
from $\xi_*$ to $\xi$,
and let $J_w(\eta)$ be the Chen integral of $d\log\mathcal L_j$ with
the convention in Eq.~\eqref{eq:gpl-chen}. For every radial
prefix define
\begin{equation}
 \mathcal R(w)=\underset{\rho\to0^+}{\operatorname{FP}_{\ell}}
 I_w\bigl(X(\tfrac12,\xi_*)\longrightarrow X(\rho,\xi_*)\bigr),
 \qquad \ell=\log\rho,\qquad \mathcal R(\varnothing)=1.
 \label{eq:tc-radial-matching-constant}
\end{equation}
The finite part is the coefficient of $\rho^0\ell^0$, with the
normalization of $\rho$ fixed by Eq.~\eqref{eq:tc-kinematics}.

At finite $\rho$, concatenate the radial path $\gamma_\rho$ with the
shape path $\eta_\rho$ at that fixed value of $\rho$. Then
\begin{equation}
 I_{(w_1,\ldots,w_n)}(\gamma_\rho\star\eta_\rho)
 =\sum_{k=0}^{n}
 I_{(w_1,\ldots,w_k)}(\gamma_\rho)
 I_{(w_{k+1},\ldots,w_n)}(\eta_\rho).
 \label{eq:tc-chen-concatenation}
\end{equation}
The shape derivatives have no $d\log\rho$ term, and their leading
forms are exactly $d\log\mathcal L_j$. Taking the finite part and
using Eq.~\eqref{eq:tc-finite-part-phase} yields
\begin{align}
 G_{11}(\xi)={}&
 \sum_{w,\gamma}q_{w,\gamma}\gamma
 \sum_{k=0}^{|w|}\mathcal R(w_1,\ldots,w_k)
 J_{(w_{k+1},\ldots,w_{|w|})}(\eta)
 +i\pi R_1(\xi)+\pi^2R_2(\xi),\notag\\
 R_2={}&V(U+V),\qquad R_1=-2(U+V)\kappa_1.
 \label{eq:tc-complete-function}
\end{align}
Equation~\eqref{eq:tc-complete-function} fixes $G_{11}$ on the chosen
branch. Within a fixed nonsingular homotopy class, its path independence
follows from that of the complete form factor.
The individual Chen words may depend on the path.
The ancillary data contain its rational coefficients, source constants,
and finite-letter and radial GPL maps. At $\xi_*$,
$U=V=\kappa_1=\kappa_2=0$, so only the empty shape suffixes contribute.

Since $\tau$ and $\sigma$ leave $\rho$ and $L$ unchanged,
the exact crossing relations in Eq.~\eqref{eq:seed} give
\begin{equation}
 G_{12}=(1-\tau)G_{11},\qquad
 G_{21}=\sigma G_{12},\qquad G_{22}=\sigma G_{11}.
 \label{eq:tc-finite-crossing}
\end{equation}
Here $\tau:(x,y,a,b)\mapsto(y,x,b,a)$ and
$\sigma:(x,y,a,b)\mapsto(x,h,c,b)$ both reverse the oriented Gram root
$d$, whereas the principal $\sqrt{\Delta_5}$ is invariant.
All roots and boundary values follow the continuation of the
complete four-point functions. Equation~\eqref{eq:tc-complete-function}
and crossing therefore fix all four finite functions, with no new
boundary constants.

\paragraph*{Relations among all nine limits.}
The ancillary files also give the complete leading-power limits of
$h_1,h_2,h_3,k_1,k_2$. With $\eta=(23)$ acting as
$(x,y,a,b,d)\mapsto(h,y,a,c,-d)$, their relations are
\begin{align}
 h_2^{\rm LP}&=\sigma h_1^{\rm LP},&
 h_3^{\rm LP}&=\eta h_1^{\rm LP},&
 \tau h_1^{\rm LP}&=h_1^{\rm LP},\notag\\
 k_1^{\rm LP}&=h_1^{\rm LP}-h_3^{\rm LP},&
 k_2^{\rm LP}&=h_2^{\rm LP}-h_3^{\rm LP}.
 \label{eq:tc-all-nine-relations}
\end{align}
Together with Eq.~\eqref{eq:seed}, these express all nine limits through
the two seeds $f_{11}^{\rm LP}$ and $h_1^{\rm LP}$, retaining their scale
logarithms and boundary constants. The files use the same pure-function
labels for the complete expressions and their symbols.

The leading-color seed has a classical-polylogarithm representation.
Defining $m=\log(abxy)$ and
\begin{equation}
 M=-L^2-mL+\beta(x,y;a,b)+9\zeta_2,
 \qquad (u,v,w)=\left(\frac{x}{1-b},\frac{y}{1-a},
                      \frac{ab}{(1-a)(1-b)}\right),
 \label{eq:tc-leading-one-loop}
\end{equation}
the BDS factorization relation of Ref.~\cite{planar} and the conversion
to our minimal-pole convention give
\begin{equation}
 h_1^{\rm LP}=\frac12M^2-2\zeta_2M
 -\zeta_3(2L+m-2i\pi)+6\zeta_4+4R_6^{(2)}(u,v,w).
 \label{eq:tc-leading-seed}
\end{equation}
Here $M$ is the one-loop finite coefficient, $\beta$ is defined in
Eq.~\eqref{eq:tc-beta}, and $R_6^{(2)}$ is the complete six-point
remainder of Ref.~\cite{classical-symbols}. Its square root is
$\sqrt{\Delta_6}=d/[(1-a)(1-b)]$, continued on the specified decay
sheet. The common $9\zeta_2$ in $M$ cancels from the ordering
differences in Eq.~\eqref{eq:tc-kappa} but is required here.
The ancillary entries explicitly expand $R_6^{(2)}$ into classical
polylogarithms. We checked their symbols exactly and their complete
values against the continued generic functions as $\rho\to0^+$.

These expressions describe
the four-point hard-function limit. For the $N^2S$ and $ND$ sectors,
extracting the two-loop hard splitting contribution additionally requires
the lower-loop products in Eq.~\eqref{eq:tc-hard-factorization}.

\subsubsection*{Boundary integrals in terms of GPLs}

On the one-dimensional curve $X(\rho,\xi_*)$, the Gram root is rational,
$\sqrt{\Delta_5}=i\rho(1-\rho)/(3\sqrt3)$.
The remaining square roots in the letters of $f_{11}$ reduce to
rational factors times
\begin{equation}
 A_\rho=\sqrt{4-\rho^2},\qquad
 B_\rho=\sqrt{(3-\rho)(1+\rho)},
 \qquad A_\rho,B_\rho>0\quad(0<\rho\leq\tfrac12).
 \label{eq:tc-radial-two-roots}
\end{equation}

No word in Eq.~\eqref{eq:tc-full-chen-sum} contains both root classes.
Each required prefix therefore contains at most one quadratic radical,
which can be rationalized separately.

On this curve, the logarithmic kernels reduce to ten distinct kernels,
represented by the letters
\begin{equation}
 \rho,\quad1-\rho,\quad2-\rho,\quad1+\rho,\quad
 \rho^2-\rho+1,\quad g_\rho,\quad
 g_{A,1},g_{A,2},g_{B,1},g_{B,2},
 \label{eq:tc-radial-kernel-basis}
\end{equation}
where
\begin{align}
 g_\rho&=-\frac{\sqrt3\rho-i\rho-\sqrt3-i}
                      {\sqrt3\rho+i\rho-\sqrt3+i},\notag\\
 g_{A,1}&=-\frac{\sqrt3A_\rho+\rho-2}
               {\sqrt3A_\rho-\rho+2},&
 g_{A,2}&=-\frac{\sqrt3A_\rho-3\rho}
               {\sqrt3A_\rho+3\rho},\notag\\
 g_{B,1}&=-\frac{\sqrt3B_\rho+3\rho-3}
                         {\sqrt3B_\rho-3\rho+3},&
 g_{B,2}&=-\frac{\sqrt3B_\rho-\rho-1}
                         {\sqrt3B_\rho+\rho+1}.
 \label{eq:tc-radial-algebraic-letters}
\end{align}
The exact integer maps from the original letters to these kernels
are supplied in the ancillary data. Constant letters give zero kernels.

Use the following separate changes of variable:
\begin{equation}
 \begin{array}{c|c|c|c|c}
 \text{sector}&\rho(u)&\text{positive radical}&u_0&u_{\rho=0}\\\hline
 A&\displaystyle\frac{4u}{1+u^2}
  &\displaystyle\frac{2(1-u^2)}{1+u^2}&4-\sqrt{15}&0\\[3pt]
 B&\displaystyle1-\frac{4u}{1+u^2}
  &\displaystyle\frac{2(1-u^2)}{1+u^2}&4-\sqrt{15}&2-\sqrt3
 \end{array}
 \label{eq:tc-radial-rationalizations}
\end{equation}
Purely rational words use the $A$ coordinate. Neither real path
contains an interior pole. Logarithmic endpoints are treated by
Eq.~\eqref{eq:tc-radial-matching-constant}.

After substitution, each letter is a rational function over
$\mathbb Q(i,\sqrt3)$, whose numerator and denominator factor into
the eight linear polynomials and two quadratics
\begin{equation}
 u,\quad u-1,\quad u-i,\quad u+i,\quad
 u\pm(2-\sqrt3),\quad u\pm(2+\sqrt3),\quad
 u^2-2(1\pm i\sqrt3)u+1.
 \label{eq:tc-gpl-polynomials}
\end{equation}
Thus the GPL pole alphabet has twelve elements,
\begin{equation}
 \mathcal A_{\rm GPL}=
 \{0,1,i,-i,\pm(2-\sqrt3),\pm(2+\sqrt3),
                  \alpha_+,\alpha_-,\bar\alpha_+,\bar\alpha_-\},
 \qquad
 \alpha_\pm=1+i\sqrt3\pm\sqrt{-3+2i\sqrt3}.
 \label{eq:tc-gpl-alphabet}
\end{equation}
The sign of the last square root only exchanges the quadratic roots.
Equation~\eqref{eq:gpl-chen-conversion}, with $t_0=u_0$, converts every
radial word into GPLs, with paths and branches inherited from the
original integral.

The endpoint finite part is normalized by $\log\rho$. The relevant
conversions from the two integration coordinates are
\begin{align}
 \log u&=\log\rho-\log4+O(\rho^2)
       &&(A\text{ sector}),\notag\\
 \log(2-\sqrt3-u)&=
 \log\rho-\log(\tfrac32+\sqrt3)+O(\rho)
       &&(B\text{ sector}).
 \label{eq:tc-gpl-finite-part-jacobians}
\end{align}
Retaining these constants in the GPL endpoint expansions fixes every
$\mathcal R(w)$ in Eq.~\eqref{eq:tc-complete-function}.

All rational coefficients and maps are exact. Source constants from
Ref.~\cite{decay-data} are retained, including weight-three and
weight-four constants supplied only as decimals with at least sixty
digits. No new boundary constants are fitted, and no analytic reduction
of these source constants to GPL periods is assumed.

An independent recursion with the rationalized GPL kernels agrees
with the original 56-letter evaluation at
$\rho=0.1,0.01,10^{-8}$ to absolute differences
$2.3\times10^{-13}$, $1.5\times10^{-12}$, and
$1.4\times10^{-10}$, respectively.

\subsubsection*{Numerical evaluation}

We combine the source constants at 85-digit working precision and integrate
Eq.~\eqref{eq:gpl-chen} in complex double precision using the DOP853
integrator in \texttt{SciPy}~\cite{scipy}. Factoring the
path polynomials before evaluation avoids cancellations in the Gram
determinant. A regular path connects $X_{\rm dec}$ to $\rho=1/2$ at
the target shape, followed by integration in $\log(2\rho)$ at that
fixed shape.

For direct integration in the 26-letter shape space, let $Y$ contain
the Chen prefixes and $f_{11}$, and let $N_\rho$ be the nilpotent
residue matrix for evolution in $\log\rho$, fixed by the valuations
$\nu_j$. Then
\begin{equation}
 Y(\rho,\xi_0)=e^{N_\rho\ell}H_0(\xi_0)+o(1),
 \qquad
 H_0(\xi_0)=\lim_{\rho\to0^+}e^{-N_\rho\ell}Y(\rho,\xi_0).
 \label{eq:tc-boundary-initial-vector}
\end{equation}
We approximate this limit at small finite $\rho$ at
$\xi_0=(1/5,1/4,1/4,1/3)$, then integrate the leading shape system.
This gives the same finite function as
Eq.~\eqref{eq:tc-complete-function}, without a separate small-$\rho$
evaluation at each target. The phase correction in
Eq.~\eqref{eq:tc-finite-part-phase} is included.
Table~\ref{tab:tc-boundary-values} gives three benchmarks at the fixed
ratios $(u,v,w)=(3/10,1/3,1/6)$ of
Eq.~\eqref{eq:tc-alph-cross-ratios}.
\begin{table}[ht]
 \centering
 \begin{tabular}{ccccc}
 \toprule
 $a$&$x$&$y$&$b$&$G_{11}$\\\midrule
 $1/5$&$9/50$&$4/15$&$2/5$&$60.74526004-25.69978517\,i$\\
 $1/4$&$1/5$&$1/4$&$1/3$&$60.47914745-25.09882891\,i$\\
 $1/3$&$9/40$&$2/9$&$1/4$&$61.46143800-23.84513246\,i$\\
 \bottomrule
 \end{tabular}
 \caption{Numerical benchmarks for $G_{11}$ in the timelike convention
 of Eq.~\eqref{eq:tc-full-logarithms}, with
 $(u,v,w)=(3/10,1/3,1/6)$. The estimated absolute uncertainty is a few
 $10^{-8}$.}
 \label{tab:tc-boundary-values}
\end{table}

At $a=1/5$ and $1/3$, evaluation of the full function at
$\rho=10^{-12}$ followed by logarithmic subtraction agrees with
direct shape integration to $2.41\times10^{-10}$ and
$7.85\times10^{-11}$, respectively. Changing shape paths and
integration tolerances shifts the values by less than
$3\times10^{-13}$. The dominant variation, about $1.4\times10^{-8}$,
comes from changing the initial value of $\rho$ from $10^{-10}$ to
$10^{-12}$. We therefore quote an empirical absolute uncertainty of
a few $10^{-8}$ at these points.

At the symmetric shape the scale coefficients vanish, giving
\begin{equation}
 G_{11}(\xi_*)\simeq-18.04863934-56.45433746\,i.
 \label{eq:tc-symmetric-numerics}
\end{equation}
This agrees, within $1.5\times10^{-9}$, with integration from $\xi_0$.
Independent finite-$\rho$ values from
\texttt{PentagonFunctions++}~\cite{decay-functions,decay-data,pentagon-cpp} at
$\rho=0.5,0.1,0.01$ differ from the Chen calculation by
$3.1\times10^{-9}$, $1.3\times10^{-8}$, and $1.24\times10^{-7}$,
respectively, in double precision. These comparisons check the
continuation and normalization.

The logarithms and alphabets apply to all four $N^0$ coefficients.
The numerical calculations and benchmarks reported here concern
$G_{11}$. The other finite functions are fixed by
Eq.~\eqref{eq:tc-finite-crossing}. Their numerical evaluation must use
the same crossing continuation.

\fi
\end{document}